\documentclass[12pt,draftcls,onecolumn] {IEEEtran}
\usepackage{cite}
\usepackage[cmex10]{amsmath}
\usepackage{pifont}
\usepackage{stmaryrd}
\usepackage{mathrsfs}
\usepackage{amsmath}
\usepackage{amssymb}
\usepackage{multirow}
\usepackage{graphicx}
\usepackage{subfigure}
\usepackage{color}
\usepackage{subeqnarray}
\usepackage{slashbox}
\usepackage{multirow}
\usepackage{cases}
\usepackage{dsfont}
\usepackage[colorlinks,
            linkcolor=black,
            anchorcolor=black,
            citecolor=black
            ]{hyperref}

\newcommand{\bs}{{\mbox{\boldmath{$s$}}}}
\newcommand{\bu}{{\mbox{\boldmath{$u$}}}}
\newcommand{\bv}{{\mbox{\boldmath{$v$}}}}

\newcommand{\bD}{{\mbox{\boldmath{$D$}}}}

\newcommand{\bS}{{\mbox{\boldmath{$S$}}}}
\newcommand{\bU}{{\mbox{\boldmath{$U$}}}}
\newcommand{\bV}{{\mbox{\boldmath{$V$}}}}
\newcommand{\bW}{{\mbox{\boldmath{$W$}}}}

\newcommand{\bgamma}{{\mbox{\boldmath{$\gamma$}}}}

\begin{document}
\title{MIMO OFDM Radar IRCI Free Range Reconstruction with Sufficient Cyclic Prefix}
\author{Xiang-Gen Xia, Tianxian Zhang, and Lingjiang Kong
\thanks{Xiang-Gen Xia is with the Department of Electrical and
Computer Engineering, University of Delaware, Newark, DE 19716, USA. Email:
xxia@ee.udel.edu. Xia's research was partially supported by the Air Force
Office of Scientific Research (AFOSR) under Grant FA9550-12-1-0055. Tianxian
Zhang  and Lingjiang Kong are with the School of Electronic Engineering,
University of Electronic Science and Technology of China, Chengdu, Sichuan,
P.R. China, 611731. Fax: +86-028-61830064, Tel: +86-028-61830768, E-mail:
tianxian.zhang@gmail.com, lingjiang.kong@gmail.com. Zhang's research was
supported by the Fundamental Research Funds for the Central Universities under
Grant ZYGX2012YB008 and by the China Scholarship Council (CSC) and was done
when he was visiting the University of Delaware, Newark, DE 19716, USA. }}

\maketitle% txzhang@udel.edu

\vspace{-0.4in}

\begin{abstract}

In this paper, we propose MIMO OFDM radar with sufficient cyclic prefix (CP),
where all OFDM pulses transmitted from different transmitters share the same
frequency band and are orthogonal to each other for every subcarrier in the
discrete frequency domain.  The orthogonality is not affected by time delays
from transmitters. Thus, our proposed MIMO OFDM radar has the same range
resolution as single transmitter radar and achieves full spatial diversity.
Orthogonal designs are used to achieve this orthogonality across the
transmitters, with which it is only needed to design OFDM pulses for the first
transmitter. We also propose a joint pulse compression and pulse coherent
integration for range reconstruction. In order to achieve the optimal SNR for
the range reconstruction, we apply the paraunitary filterbank theory to design
the OFDM pulses. We then propose a modified iterative clipping and filtering
(MICF) algorithm for the designs of OFDM pulses jointly, when other important
factors, such as peak-to-average power ratio (PAPR) in time domain, are also
considered. With our proposed MIMO OFDM radar, there is no interference for the
range reconstruction not only across the transmitters but also across the range
cells in a swath called inter-range-cell interference (IRCI) free that is
similar to our previously proposed CP based OFDM radar for single transmitter.
Simulations are  presented to illustrate our proposed theory and show that the
CP based MIMO OFDM radar outperforms the existing frequency-band shared MIMO
radar with polyphase codes and also frequency division MIMO radar.

\end{abstract}

\clearpage

\maketitle

\section{Introduction}\label{Introduction}
Multiple-input multiple-output (MIMO) concept using multiple transmit and
receive antennas has %firstly
been intensively investigated in the last decades in wireless communications to
collect spatial diversity, see, for example, \cite{TseFundamentals2005,
BiglieriMIMO2007}. In recent years, this concept has been introduced to the
radar applications \cite{HaimovichIEEESPMMIMOWidely2008116,
JianIEEESPMMIMOColocated2007106, JianMIMO2008}, which is named as ``MIMO
radar.'' Unlike the traditional mono-static radar or phased-array radar, MIMO
radar systems employ multiple transmitters, multiple receivers and multiple
orthogonal signals, and can provide more degrees of freedom for the design of a
radar system as well as more advantages for radar signal processing. According
to the configuration of antennas/transmitters, MIMO radar systems can be
divided into two types, namely statistical MIMO radar and colocated MIMO radar.
For statistical MIMO radar, the transmitters and receivers are widely
separated, then, a target can be observed from different spatial aspects,
resulting in spatial diversity and performance improvements of target detection
\cite{HaimovichIEEESPMMIMOWidely2008116},  synthetic aperture radar
(SAR) applications \cite{JungHyoKimNovelOFDMChirpIEEEGRSL2013568},
and direction of arrival estimation \cite{WuMIMOOFDMIETRSN201028,
SitDirectionOfArrivalMIMOOFDMIEEERadarCon20120226}. For
colocated MIMO radar, the transmitters and receivers are located closely
enough. By exploiting waveform diversity, colocated MIMO radar can improve the
flexibility for transmit beam design \cite{JianIEEESPMMIMOColocated2007106,
JianMIMO2008},
and low-grazing angle target
racking \cite{SenTSPOFDMMIMO20103152}.

The above advantages of MIMO radar systems are achieved under the assumption
that the transmitted signals are orthogonal to each other in time domain
despite their arbitrary time delays. It is well known that this assumption can
hold only when the frequency bands of all the transmitted signals do not
overlap  each other \cite{SanAntonioMIMOAFIEEEJSTSP2007167}. Then, the signals
of different transmitter and receiver pairs can be independently processed and
the spatial diversity can be obtained. This MIMO radar system can be denoted as
``frequency division MIMO radar'' system, which requires a relatively wide
frequency band, since each transmitter occupies a unique frequency band.
Therefore, the frequency spectrum efficiency is low, especially, for a high
range resolution radar system. In other words, the spatial diversity advantage
of frequency division MIMO radar systems is built upon the sacrifice of the
range resolution. To increase the spectrum efficiency or  the range resolution
of frequency division MIMO radar systems, there have been many works on
investigating  ``frequency-band shared MIMO radar'' systems through the design
of time domain orthogonal codes/sequences and/or waveforms, which contain not
only good autocorrelation but also good cross-correlation properties
\cite{SomainiUBinarySequencesIEEETAES19751226,
HaiDengSynthesisOfBinarySequencesIEEETAES199698,
HaiDengPolyphaseCodeIEEETSP20043126, KhanHAOptimizingPolyphaseIEEESPL2006589,
HaoHeDesigningUnimodularSequenceIEEETSP20094391,
XiuSengReducingIEEETSP20104213, LeiXuZeroCorrelationZoneIEEETAES20122100,
JinYComplementaryBasedIETRSN2013371}. However, the design of binary sequences
\cite{SomainiUBinarySequencesIEEETAES19751226,
HaiDengSynthesisOfBinarySequencesIEEETAES199698}, polyphase sequences
\cite{HaiDengPolyphaseCodeIEEETSP20043126,
KhanHAOptimizingPolyphaseIEEESPL2006589}, unimodular sequence sets
\cite{HaoHeDesigningUnimodularSequenceIEEETSP20094391} or chaotic phase codes
\cite{JinYComplementaryBasedIETRSN2013371} can only somewhat mitigate waveform
cross-correlation effects or reduce the sidelobes of autocorrelation function.
The cross correlations between the delayed time domain waveforms from different
transmitters can not be zero and thus cause interference among transmitters.
This limits the collection of the spatial diversity. Therefore, the performance
of MIMO radar systems will still be degraded by using the existing
designed waveforms. % with non-ideal autocorrelations and cross-correlations,
%especially, for the case of large amount of transmitters.

To deal with the sidelobe issues from the non-ideal autocorrelations across the
range cells in the conventional SAR systems, in \cite{TxzOFDMSAR,
TxzAPLOFDMSAR} we have proposed a sufficient cyclic prefix (CP) based
orthogonal frequency division multiplexing (OFDM) SAR imaging for single
transmitter radar systems. By using a sufficient CP, %an ideal autocorrelation property with
zero range sidelobes and inter-range-cell interference (IRCI) free range
reconstruction can be achieved, which provides an opportunity for high
resolution range reconstruction. As it has been explained in \cite{TxzOFDMSAR},
the major differences between our proposed CP based OFDM SAR and the existing
OFDM SAR systems are in two aspects. One is that a sufficiently long CP is used
at the transmitter and the CP should be as long as possible when the number of
range cells in a swath is large. The other is the SAR imaging algorithm at the
receiver,
 which is not  the matched filter receiver
 by simply treating the CP based OFDM signals as
radar waveforms as what is done in the existing OFDM radar systems. With these
two differences, the key feature of an OFDM system in  communications
applications of converting an intersymbol interference (ISI) channel to
multiple  ISI free subchannels is analogously obtained in our proposed CP based
OFDM SAR imaging as IRCI free range reconstruction among range cells in a
swath.

%However, for the range reconstruction of frequency-band shared MIMO radar systems,
%the study of how to eliminate the influences between different transmitters and
%achieve spatial diversity is still an open problem in the available literature.

In this paper, we consider a frequency-band shared statistical
 MIMO radar range
reconstruction using OFDM signals with sufficient CP  by generalizing the CP
based OFDM SAR imaging from single transmitter and receiver to multiple
transmitter and receiver  radar systems called ``MIMO OFDM radar.'' With our
newly proposed CP based MIMO OFDM radar, all the signal waveforms from all the
transmitters have the same frequency band and thus the range resolution is not
sacrificed and the same as the single transmitter radar. Furthermore, their
arbitrarily time delayed versions are still orthogonal
for every subcarrier in the discrete  frequency domain and
therefore, the spatial diversity from all the transmitters can be collected the
same as the frequency division MIMO radar. In addition to the two differences
mentioned above for single transmitter and receiver CP based OFDM radar systems
with   the existing OFDM radar systems, the orthogonality in the time
domain under arbitrarily time delays between different transmitters have not
been considered in most of the existing MIMO OFDM radar systems
\cite{JungHyoKimNovelOFDMChirpIEEEGRSL2013568, %WuMIMOOFDMIETRSN201028,
SitDirectionOfArrivalMIMOOFDMIEEERadarCon20120226, SenTSPOFDMMIMO20103152}
where IRCI exists not only  among  range cells in a swath but also among the
transmitters.
Although it is considered in \cite{WuMIMOOFDMIETRSN201028},
IRCI is not the focus.
In this paper, IRCI free is achieved among both range cells in a
swath and all the transmitters.
%WuMIMOOFDMIETRSN201028,

We first formulate the problem  and describe the MIMO OFDM radar signal model
by considering the feature of sufficient CP based OFDM pulses, where the CP
part takes all zero values. Using the properties of frequency domain orthogonal
OFDM pulses for every subcarrier
between different transmitters, we then derive a MIMO OFDM radar
range reconstruction algorithm, which includes the joint processing of pulse
compression and pulse coherent integration. We also analyze the change of noise
power in every step of the range reconstruction and evaluate the possible
signal-to-noise ratio (SNR) degradation caused by the range reconstruction. We
then propose the design criterion for the multiple OFDM pulses used at
transmitters.

The orthogonality for every subcarrier  in the discrete
frequency domain among the OFDM waveforms for all the
transmitters is done by employing the theory of orthogonal designs
\cite{AlamoutiASimpleIEEEJSAC19981451,TarokhSpaceTimeBlockIEEETIT19991456,
SuWeiSpaceTimeWPC20031, XueBinOnTheNonexistenceIEEETIT20032984,
HaiQuanUpperBoundsIEEETIT20032788,XueBinOrthogonalDesignsIEEETIT20032468,
SuWeiASystematicDesignIEEECL2004380,KejieClosedFormDesignsIEEETIT20054340} that
has been used as orthogonal space-time codes in MIMO wireless communications
\cite{TseFundamentals2005, BiglieriMIMO2007,
AlamoutiASimpleIEEEJSAC19981451,TarokhSpaceTimeBlockIEEETIT19991456,
SuWeiSpaceTimeWPC20031, XueBinOnTheNonexistenceIEEETIT20032984,
HaiQuanUpperBoundsIEEETIT20032788,XueBinOrthogonalDesignsIEEETIT20032468,
SuWeiASystematicDesignIEEECL2004380,KejieClosedFormDesignsIEEETIT20054340}. To
achieve the optimal SNR after the range reconstruction,
 we propose a joint multiple OFDM pulse design method with a closed-form
solution by using paraunitary filterbank theory
\cite{PPVaidyanathanMultirate1993, XiangGenXiaMultirateFilterbanks}.
%The
%designed OFDM pulses can guarantee the frequency domain orthogonality and
%without any SNR degradation within the range reconstruction.
With the paraunitary filterbank theory in the design of the MIMO OFDM waveforms,
although the SNR after the range reconstruction is maximized,
it is not easy to search for the sequences to generate the MIMO OFDM waveforms
so that their
peak-to-average power
ratio (PAPR) is low, while a low PAPR is important in radar
practical implementations. By considering the trade-off between the
PAPR and the SNR degradation within the range reconstruction, we propose a
modified iterative clipping and filtering (MICF) joint OFDM pulse design
method, which can obtain OFDM pulses with low PAPRs and an acceptable SNR
degradation. We then present some simulations to demonstrate the performance of
the proposed MICF joint OFDM pulse design method. By comparing with the
frequency-band shared MIMO radar using polyphase code waveforms and frequency
division MIMO radar using linear frequency modulated (LFM) waveforms, we
present some simulations to illustrate the performance advantage of the
proposed MIMO OFDM radar range reconstruction method. We find that, with the
designed OFDM pulses from our proposed MICF method, our proposed CP based MIMO
OFDM radar can obtain the range reconstruction without any interference
 between
different transmitters and achieve the full spatial diversity from all the
transmitters and receivers. Meanwhile, it can still maintain the advantage of
IRCI free range reconstruction with insignificant SNR degradation and
completely avoid the energy redundancy in the case when there are only a limited number of
range cells of interest.
Note that constant orthogonal/unitary
 matrices for every subcarrier in the discrete frequency domain across transmitters and waveforms have been constructed in
\cite{WuMIMOOFDMIETRSN201028}
 where only a few parameters are used and may limit the waveform designs with other desired properties, such as those discussed above.

The remainder of this paper is organized as follows. In Section
\ref{Problem_For}, we establish the CP based MIMO OFDM radar signal model and
describe the problem of interest. In Section \ref{Range_Reconstruction}, we
propose CP based MIMO OFDM radar range reconstruction. In Section
\ref{OFDM_Design}, we propose two new arbitrary length OFDM sequence design
methods. In Section \ref{Simulation}, we show some simulation results. Finally,
in Section \ref{Conclusion}, we conclude this paper.

\section{CP Based MIMO OFDM Radar Signal Model and Problem Formulation}\label{Problem_For}
\begin{figure}[t]
\begin{center}
\includegraphics[width=0.6\columnwidth,draft=false]{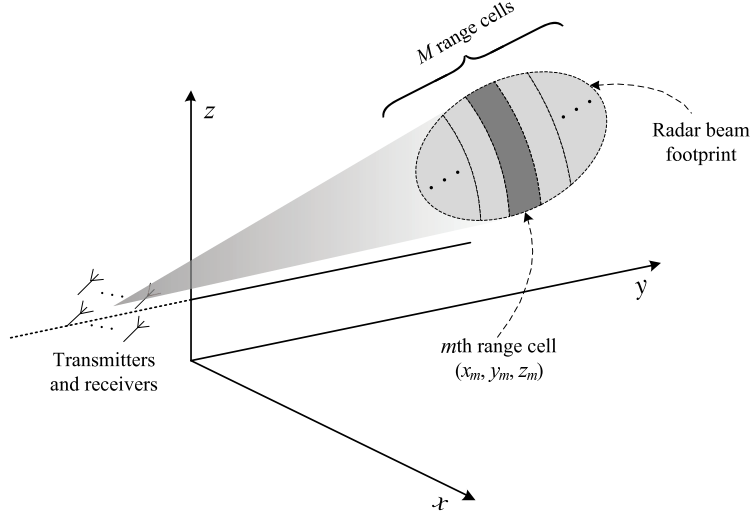}
\end{center}
\caption{MIMO OFDM radar geometry.}\label{geometry}
\end{figure}

Consider a MIMO radar system with $\mathbb{T}$ transmitters and $\mathbb{R}$
receivers, as shown in Fig. \ref{geometry}. All the antennas of a MIMO radar
system we consider in this paper are located in a fixed area, and the antennas
are not as close to each other as colocated MIMO radars
\cite{JianIEEESPMMIMOColocated2007106,JianMIMO2008}. The instantaneous
coordinate of the $\alpha$th transmitter and the $\beta$th receiver
are, respectively,
$\left(x_{\alpha},y_{\alpha},z_{\alpha}\right),\
\alpha=1,\ldots,\mathbb{T}$, and
$\left(x_{\beta},y_{\beta},z_{\beta}\right),\
\beta=1,\ldots,\mathbb{R}$, where $z_{\alpha}$ and
$z_{\beta}$ are the altitudes of the corresponding antennas. After the
demodulation to baseband, the complex envelope of the received signal observed
at the $\beta$th receiver due to a transmission from the $\alpha$th
transmitter and reflection from the far field scatterers in the $m$th range
cell with instantaneous coordinate $\left(x_m,y_m,z_m\right)$ (and excluding
noise) is given by
\begin{equation}\label{urtm}
u_{\beta,\alpha,m}(t)=g_{\beta,\alpha,m}\textrm{exp}
\left\{-j2\pi
f_c\left[\tau_{\alpha,m}+\tau_{\beta,m}\right]\right\}s_{\alpha}
\left(t-\tau_{\alpha,m}-\tau_{\beta,m}\right),
\end{equation}
where $s_{\alpha}\left(t\right)$ is a transmitted signal of the
$\alpha$th transmitter, $f_c$ is the carrier frequency,
$g_{\beta,\alpha,m}$ is the radar cross section (RCS) coefficient
caused from the scatterers in the $m$th range cell within the radar main beam
footprint and related to the $\alpha$th transmitter and the
$\beta$th receiver. We assume that the main beam footprints of each
receiver are overlapped together and included in the footprints of the
transmitters. $\tau_{\alpha,m}=\frac{R_{\alpha,m}}{c}$ is the signal
propagation time delay between the $\alpha$th transmitter and the $m$th
range cell, and similarly, $\tau_{\beta,m}=\frac{R_{\beta,m}}{c}$
is the signal propagation time delay between the $m$th range cell and the
$\beta$th receiver, where $c$ is the speed of light,
$R_{\alpha,m}=\sqrt{\left({x}_m-x_\alpha\right)^2
+\left({y}_m-y_\alpha\right)^2+\left(z_m-z_{\alpha}\right)^2}$ and
$R_{\beta,m}=\sqrt{\left({x}_m-x_\beta\right)^2
+\left({y}_m-y_\beta\right)^2+\left(z_m-z_{\beta}\right)^2}$ are,
respectively, the slant range between the $\alpha$th transmitter and the
$m$th range cell and the slant range between the $m$th range cell and the
$\beta$th receiver.

At the receiver, to a transmitted signal with bandwidth $B$, the received
signal is sampled by the A/D converter with sampling interval length
$T_s=\frac{1}{B}$ and the range resolution is
$\rho=\frac{c}{2B}=\frac{c}{2}T_s$. Assume that the width for the radar
footprints in the range direction is $R_w$. Then, a range profile can be
divided into $M=\frac{R_w}{\rho}$ range cells as in Fig. \ref{geometry} that is
determined by the radar system. From the far field assumption, as we have
discussed in \cite{TxzOFDMSAR}, we can obtain
\begin{subequations}\label{RtmRrm}
\begin{align}
R_{\alpha,m}&=R_{\alpha,0}+m\rho,\ m=0,1,\ldots,M-1,\\
R_{\beta,m}&=R_{\beta,0}+m\rho,\ m=0,1,\ldots,M-1.
\end{align}\end{subequations}
Then, the signal propagation time delay between the $\alpha$th transmitter
and the $\beta$th receiver can be denoted by
\begin{equation}\label{tautmtaurm}
  \tau_{\alpha,m}+\tau_{\beta,m}=\tau_{\beta,\alpha,0}+mT_s,
\end{equation}
where
\begin{equation}
  \tau_{\beta,\alpha,0}=\frac{R_{\alpha,0}+R_{\beta,0}}{c}.
\end{equation}
In radar applications, there are usually more than one scatterers within a
range cell, and each scatterer owns its unique delay and phase. However, for a
given range resolution (or signal bandwidth), a radar is not able to
distinguish these individual scatterers, and the responses of all these
scatterers are summarized as the response of one range cell with a single delay
and phase in the receiver. Thus, each range cell can be treated as one
point-like target. This kind of model is reasonable and commonly used in the
existing radar applications \cite{skolnik2001Introduction}.

Let $\tau_{min}$ be the minimal signal propagation time delay between all the
transmitter and receiver pairs through the nearest ($m=0$) range cell. And
$\tau_{min}$ is defined as
\begin{equation}
\tau_{min}=\underset{\scriptstyle{\beta=1,\ldots,\mathbb{R}}
\atop\scriptstyle{\alpha=1,\ldots,\mathbb{T}}}\min\left\{\tau_{\beta,\alpha,0}\right\}.
\end{equation}
By arranging the antennas, the time delays between different transmitter and
receiver pairs can approximately satisfy the relationship
\begin{equation}\label{MrtTs}
{\eta}_{\beta,\alpha}=\frac{\tau_{\beta,\alpha,0}-\tau_{min}}{T_s},
\end{equation}
where ${\eta}_{\beta,\alpha}\in \mathbb{N}$. The maximal relative
time delay difference among all the transmitter and receiver pairs is
${\eta}_{max}T_s$, and
\begin{equation}\label{barMmax}
{\eta}_{max}=\underset{\beta,\alpha}
\max\left\{{\eta}_{\beta,\alpha}\right\}.
\end{equation}
We remark that the values of ${\eta}_{\beta,\alpha}$ may slightly
change, when a radar scans the radar surveillance area with different azimuth angle.
But, in practice, considering the far field assumption,
${\eta}_{\beta,\alpha}$ is constant with in a consecutive radar scan
sector. Thus, the radar surveillance area can be divided into different radar
scan sectors with different precalculated values of
${\eta}_{\beta,\alpha}$. Also,  parameter ${\eta}_{max}$
is determined by the system configuration and may be estimated in priori,
and it will be used for the MIMO OFDM pulse designs later.

In most of the MIMO radar literatures, it is assumed that the transmitted
signals are orthogonal to each other and even when there are
 different time delays among these signals,
i.e., $\int
s_{\alpha}\left(t\right)s_{\tilde{\alpha}}\left(t-\tau\right)^*dt=0$
for $\alpha\neq \tilde{\alpha}$, and arbitrary time delay $\tau$ of
interest, where $(\cdot)^*$ denotes the complex conjugate, or it is assumed that
there are no different time
delays among the transmitted signals from multiple transmitters. However, in
practice, this is  generally not possible
\cite{SanAntonioMIMOAFIEEEJSTSP2007167}, unless the frequency bands of all the
transmitted signals do not overlap with each other, which then leads to
frequency division MIMO radar and will either reduce the range resolution or
not be able to collect the transmitter spatial diversity as we have mentioned
in Introduction. As will shall see later, in this paper, these two assumptions
will not be needed with our proposed MIMO OFDM radar.

In this paper, we consider that there are $P$ coherent pulses in a radar
coherent processing interval (CPI) (as we shall see later
that some of these $P$ pulses may be all zero-valued). Each non-zero-valued
pulse is an OFDM signal with $N$
subcarriers and  a bandwidth of $B$ Hz. Let
$\bS_\alpha^{(p)}=\left[S_{\alpha,0}^{(p)},S_{\alpha,1}^{(p)},\ldots,
S_{\alpha,N-1}^{(p)}\right]^T$ represent the complex weights transmitted
over the subcarriers of the $p$th OFDM pulse and the
 $\alpha$th transmitter,
where $p=0,1,\ldots,P-1$, and $(\cdot)^T$ denotes the transpose. For
convenience, we normalize the total transmitted energy within a CPI to $1$, and
assume the energy of each transmitted pulse is the same, i.e.,
$\sum\limits_{k=0}^{N-1}\left|S_{\alpha,k}^{(p)}\right|^2=\frac{1}{\mathbb{T}P_0}$
for all non-zero-valued
pulses where $P_0$ is the
number of non-zero-valued pulses. All the transmitted signals share the same frequency band.
Then, a discrete time OFDM signal is the inverse fast Fourier transform (IFFT)
of the vector $\bS_\alpha^{(p)}$ and the corresponding time domain OFDM
signal is
\begin{equation}\label{OFDM}
s_\alpha^{(p)}(t)=\frac{1}{\sqrt{N}}\sum_{k=0}^{N-1}S_{\alpha,k}^{(p)}
\textrm{exp}\left\{j2\pi k\Delta ft\right\},\ t\in
\left[pT_r,pT_r+T+T_{GI}\right],
\end{equation}
where $\Delta f=\frac{B}{N}=\frac{1}{T}$ is the subcarrier spacing, $T_r$ is
the time interval between two consecutive pulses with in a CPI.
$\left[pT_r,pT_r+T_{GI}\right)$ is the time duration of the guard interval that
corresponds to the CP in the discrete time domain as we shall see later in more
details and its length $T_{GI}$ will be specified later as well,
$T$ is the length
of the OFDM signal excluding CP. Due to the periodicity of the exponential
function $\textrm{exp}(\cdot)$ in (\ref{OFDM}), the tail part of
$s_\alpha^{(p)}(t)$ for $t$ in $\left(pT_r,pT_r+T+T_{GI}\right]$ is the
same as the head part of $s_\alpha^{(p)}(t)$ for $t$ in
$\left[pT_r,pT_r+T_{GI}\right)$. Note that in the above
transmission, the CP is added to each pulse $s_{\alpha}^{(p)}(t)$.

Then, the complex envelope of the received signal in the $\beta$th
receiver due to the $p$th transmitted pulses of all the transmitters and the
reflection from all range cells within the main beam footprint can be written
as
\begin{equation}\label{urOFDM}\begin{split}
u_\beta^{(p)}(t)&=\frac{1}{\sqrt{N}}\sum\limits_{\alpha=1}^{\mathbb{T}}
\sum\limits_{m=1}^{M-1}g_{\beta,\alpha,m}\textrm{exp} \left\{-j2\pi
f_c\left[\tau_{\alpha,m}+\tau_{\beta,m}\right]\right\}\\
&\times\sum\limits_{k=0}\limits^{N-1}S_{\alpha,k}^{(p)}
\textrm{exp}\left\{\frac{j2\pi k}{T}
\left[t-\tau_{\alpha,m}-\tau_{\beta,m}\right]\right\}+w_{\beta}^{(p)}(t),
\end{split}\end{equation}
where $w_{\beta}^{(p)}(t)$ represents the noise. For convenience, in this
paper, we assume the RCS coefficients $g_{\beta,\alpha,m}$ are
constant within a CPI, and it can be generalized to the case of maneuvering
targets similar to what is done in the literature.

In our MIMO radar applications, the values of time delays
$\tau_{\beta,\alpha,0}$ are different from one transmitter and
receiver pair to another pair, which depend on the relative locations of
antennas. All the received signals due to $\mathbb{T}$ transmitters and
reflections from each range cell will overlap together at the receiver and can
not be separated in general. Thus, the interferences will occur including different range
cells and different transmitted signals from the transmitters and result in
IRCI. Notice that, to one range cell, each transmitter and receiver pair can be
regarded as one path of communications, and, to one transmitter and receiver
pair each range cell can also be regarded as one path of communications as
analyzed in \cite{TxzOFDMSAR}. Comparing with the main path that we define as
the shortest path, the longest time delay among all the paths is
$\left({\eta}_{max}+M-1\right)T_s$. As we have mentioned in \cite{TxzOFDMSAR},
to eliminate the interference between different transmitted signals and IRCI,
similar to OFDM systems in communications, the time duration of guard interval
should be at least $\left({\eta}_{max}+M-1\right)T_s$. For convenience, we use
CP length ${\eta}_{max}+M-1$ in this paper, i.e., a CP of length
${\eta}_{max}+M-1$ is added at the beginning of an OFDM pulse, and then the
guard interval length $T_{GI}$ in the analog transmission signal is
$T_{GI}=\left({\eta}_{max}+M-1\right)T_s$. Notice that $T=NT_s$, so the time
duration of an OFDM pulse is $T_o=T+T_{GI}=\left(N+{\eta}_{max}+M-1\right)T_s$.
To completely avoid the range
interference between different transmitted signals and
range cells, the number, $N$, of the OFDM signal subcarriers should satisfy
$N\geq{\eta}_{max}+M$ as we have analyzed in \cite{TxzOFDMSAR}
and will be seen in more details later, and it is also
well understood in communications applications \cite{prasad2004ofdm}.

\section{CP Based MIMO OFDM Radar Range Reconstruction}\label{Range_Reconstruction}
This section is on the MIMO radar range reconstruction that includes the joint
processing of pulse compression and pulse coherent integration. Going back to
(\ref{urOFDM}), for the $p$th pulse, let the sampling at all receivers be
aligned with the start of the received signals after $pT_r+\tau_{min}$ seconds
of the transmitted pulses, where we recall that $T_r$ is the time interval
between two consecutive pulses. Combining with (\ref{tautmtaurm}),
(\ref{MrtTs}) and (\ref{urOFDM}), $u_\beta^{(p)}(t)$ can be converted to
the discrete time linear convolution of the transmitted sequence with the
weighting RCS coefficients $d_{\beta,\alpha,m}$ after the sampling
$t=pT_r+\tau_{min}+iT_s$ and the received sequence can be written as
\begin{equation}\label{tildeuri}
  \tilde{u}_{\beta,i}^{(p)}=\sum_{\alpha=1}^{\mathbb{T}}\sum_{m=0}^{M-1}
  d_{\beta,\alpha,m}s_{\alpha,i-m-{\eta}_{\beta,\alpha}}^{(p)}+
  \tilde{w}_{\beta,i}^{(p)},\ i=0,1,\ldots,N+2({\eta}_{max}+M)-3,
\end{equation}
where
\begin{equation}\label{drlm}
d_{\beta,\alpha,m}=g_{\beta,\alpha,m}\textrm{exp}
\left\{-j2\pi
f_c\left[\tau_{\alpha,m}+\tau_{\beta,m}\right]\right\},\\
\end{equation}
in which $2\pi f_c\left[\tau_{\alpha,m}+\tau_{\beta,m}\right]$ in
the exponential is known and related to the target location\footnote{Notice
that the values of $j2\pi f_c\tau_{\alpha,m}$ and $j2\pi
f_c\tau_{\beta,m}$ form the transmitter steering vector and receiver
steering vector \cite{FishlerIEEETSPSpatialDiversityMIMO2006823}, respectively,
which are often assumed known.}, and $s_{\alpha,i}^{(p)}$ is the complex
envelope of the OFDM pulse in (\ref{OFDM}) with time duration $t\in
\left[pT_r,pT_r+T+T_{GI}\right]$ for $T=NT_s$ and
$T_{GI}=({\eta}_{max}+M-1)T_s$. In (\ref{tildeuri}),
$\tilde{w}_{\beta,i}^{(p)}$ is the noise. After sampling at
$t=pT_r+iT_s$, (\ref{OFDM}) can be recast as:
\begin{equation}\label{sti}
s_{\alpha,i}^{(p)}=s_{\alpha}^{(p)}(iT_s)=\frac{1}{\sqrt{N}}\sum_{k=0}^{N-1}
S_{\alpha,k}^{(p)}\textrm{exp}\left\{\frac{j2\pi ki}{N}\right\},\
i=0,1,\ldots,N+{\eta}_{max}+M-2,
\end{equation}
and $s_{\alpha,i}^{(p)}=0$ if $i<0$ or $i>N+{\eta}_{max}+M-2$.

When the sequence
$\tilde{\bu}_\beta=\left[\tilde{u}_{\beta,0},\tilde{u}_{\beta,1},
\ldots,\tilde{u}_{\beta,N+2({\eta}_{max}+M)-3}\right]^T$ in
(\ref{tildeuri}) is received, the first and the last ${\eta}_{max}+M-1$ samples
are removed as \cite{TxzOFDMSAR}, and then, we obtain
\begin{equation}\label{urn}
  {u}_{\beta,n}^{(p)}=\sum_{\alpha=1}^{\mathbb{T}}\sum_{m=0}^{M-1}
  d_{\beta,\alpha,m}s_{\alpha,n+{\eta}_{max}+M-m-
  {\eta}_{\beta,\alpha}-1}^{(p)}+
  {w}_{\beta,n}^{(p)},\ n=0,1,\ldots,N-1.
\end{equation}

The OFDM demodulator then performs the $N$-point fast Fourier transform (FFT)
on the vector ${\bu}_{\beta}^{(p)}=
\left[{u}_{\beta,0}^{(p)},\ldots,{u}_{\beta,N-1}^{(p)}\right]^T$,
and  obtains ${\bU}_{\beta}^{(p)}=
\left[{U}_{\beta,0}^{(p)},\ldots,{U}_{\beta,N-1}^{(p)}\right]^T$,
where ${U}_{\beta,k}^{(p)}$ can be denoted as
\begin{equation}\label{Urk}
 {U}_{\beta,k}^{(p)}=\bD_{\beta,k}\bar{\bS}_{k}^{(p)}+W_{\beta,k}^{(p)},\
 k=0,1,\ldots,N-1,
\end{equation}
where $\bar{\bS}_{k}^{(p)}=
\left[{S}_{1,k}^{(p)},\ldots,{S}_{\mathbb{T},k}^{(p)}\right]^T$ is a
$\mathbb{T}\times 1$ column vector. $W_{\beta,k}^{(p)}$ is the FFT of
noise, and $\bD_{\beta,k}=\left[{D}_{\beta,1,k},\ldots,
{D}_{\beta,\mathbb{T},k}\right]$ with
\begin{equation}\label{Drtk}
{D}_{\beta,\alpha,k}=\sum_{m=0}^{M-1}d_{\beta,\alpha,m}
\textrm{exp}\left\{\frac{j2\pi
k\left({\eta}_{max}+M-{\eta}_{\beta,\alpha}-1\right)}{N}\right\}
\textrm{exp}\left\{\frac{-j2\pi km}{N}\right\},\ k=0,1,\ldots,N-1,
\end{equation}
where, $d_{\beta,\alpha,m}$ is the weighting RCS coefficient from
the $\alpha$th transmitter, the $m$th range cell, and the $\beta$th
receiver.

From the constant assumption of $g_{\beta,\alpha,m}$ within a CPI,
for given $\beta$, $\alpha$ and $m$, the values of
$d_{\beta,\alpha,m}$ in (\ref{drlm}) and
${D}_{\beta,\alpha,k}$ in (\ref{Drtk}) are also constant within a
CPI. Combining all the received signals of $\mathbb{R}$ receivers and $P$
pulses within a CPI, we can obtain the following matrix representation:
\begin{equation}\label{bUk}
\mathbf{U}_k=\mathbf{D}_k\mathbf{S}_k+\mathbf{W}_k,\ k=0,1,\ldots,N-1,
\end{equation}
where $\mathbf{U}_k=\left[{\bU}_k^{({0})},{\bU}_k^{({1})},
\ldots,{\bU}_k^{({P-1})}\right]$ is a ${\mathbb{R}\times {P}}$ matrix,
${\bU}_k^{({p})}=\left[{U}_{1,k}^{({p})},
{U}_{2,k}^{({p})},\ldots, {U}_{\mathbb{R},k}^{({p})}\right]^T$ is a
$\mathbb{R}\times 1$ column vector for $0\leq p\leq P-1$.
\begin{equation}\label{bfSk}
\mathbf{S}_k\triangleq\left[\bar{\bS}_k^{({0})},\bar{\bS}_k^{({1})},
\ldots,\bar{\bS}_k^{({P-1})}\right]=
  \begin{bmatrix}
  {S}_{1,k}^{(0)} & {S}_{1,k}^{(1)} &\cdots &{S}_{1,k}^{(P-1)}\\
  {S}_{2,k}^{(0)} & {S}_{2,k}^{(1)} &\cdots &{S}_{2,k}^{(P-1)}\\
  \vdots  & \vdots   & \ddots  &\vdots \\
  {S}_{\mathbb{T},k}^{(0)}  & {S}_{\mathbb{T},k}^{(1)}  &\cdots &{S}_{\mathbb{T},k}^{(P-1)}
  \end{bmatrix}
\end{equation}
is a ${\mathbb{T}\times P}$ matrix.
$\mathbf{W}_k=\left[\bW_k^{({0})},\bW_k^{({1})}, \ldots,\bW_k^{({P-1})}\right]$
is a ${\mathbb{R}\times {P}}$ matrix,
$\bW_k^{({p})}=\left[{W}_{1,k}^{({p})},
{W}_{2,k}^{({p})},\right.$ $\ldots,$ $\left. {W}_{\mathbb{R},k}^{({p})}\right]^T$ is a
$\mathbb{R}\times 1$ column vector. And
\begin{equation}\label{D_matrix_k}
  \mathbf{D}_k=
  \begin{bmatrix}
  {D}_{1,1,k} & {D}_{1,2,k} &\cdots &{D}_{1,\mathbb{T},k}\\
  {D}_{2,1,k} & {D}_{2,2,k} &\cdots &{D}_{2,\mathbb{T},k}\\
  \vdots  & \vdots   & \ddots  &\vdots \\
  {D}_{\mathbb{R},1,k} & {D}_{\mathbb{R},2,k} &\cdots &{D}_{\mathbb{R},\mathbb{T},k}
  \end{bmatrix}
\end{equation}
is a ${\mathbb{R}\times\mathbb{T}}$ matrix.

By assuming $P\geq{\mathbb{T}}$, we can construct such a $\mathbb{T}\times{P}$
matrix $\mathbf{S}_k$ to guarantee
$\mathbf{S}_k\mathbf{S}_k^{+}=\mathbf{I}_{\mathbb{T}}$ for all $k$, where
$\mathbf{I}_{\mathbb{T}}$ is the $\mathbb{T}\times\mathbb{T}$ identity matrix,
$\mathbf{S}_k^{+}=\mathbf{S}_k^\dag\left(\mathbf{S}_k\mathbf{S}_k^\dag\right)^{-1}
\in\mathbb{C}^{P\times\mathbb{T}}$ is the Penrose-Moore pseudo-inverse of
$\mathbf{S}_k$, and $(\cdot)^\dag$ denotes the conjugate transpose. Note that
as long as matrix $\mathbf{S}_k$ has full row rank, i.e.,
$P\times 1$ weight vectors in the
$P$ OFDM waveforms from all transmitters are linearly independent on every subcarrier $k$, property
$\mathbf{S}_k\mathbf{S}_k^{+}=\mathbf{I}_{\mathbb{T}}$ is satisfied.

Then, the estimate of $\mathbf{D}_k$ in (\ref{bUk}) is
\begin{equation}\label{hatbDk}
\hat{\mathbf{D}}_k=\mathbf{U}_k\mathbf{S}_k^{+}=\mathbf{D}_k+\overline{\mathbf{W}}_k,
\end{equation}
where $\overline{\mathbf{W}}_k=\mathbf{W}_k\mathbf{S}_k^{+}$ denotes the new
noise matrix. One can see from the above estimate that the new noise matrix is
obtained by multiplying the inverse of matrix $\mathbf{S}_k$ to the original
noise matrix $\mathbf{W}_k$ for each subcarrier index $k$. Clearly, in order
not to enhance the noise, it is desired that the matrix $\mathbf{S}_k$ is
unitary, which is similar to the MIMO OFDM channel estimation in wireless
communications \cite{TseFundamentals2005, BiglieriMIMO2007, prasad2004ofdm}.
Since $\mathbf{S}_k$ is a flat matrix in general, in what follows we require
that the row vectors of $\mathbf{S}_k$ are orthogonal each other and have the
same norm called flat unitary matrix, i.e.,
$\mathbf{S}_k\mathbf{S}_k^{\dagger}=\mathbf{I}_{\mathbb{T}}$. This means that
the weight vectors at every subcarrier  $k$
 in the OFDM waveforms transmitted through $\mathbb{T}$
transmitters are orthogonal each other among different transmitters, i.e., the
discrete versions in  frequency domain are orthogonal each other for every
subcarrier,
which still holds when there are time delays  among the corresponding
 waveforms in time domain,
although
the delayed  waveforms may not be orthogonal in time domain.
{\em This property is fundamentally different from most of the existing MIMO radars including
the exisiting MIMO OFDM radars}.

According to (\ref{Drtk}),  vector
$\bD_{\beta,\alpha}=\left[D_{\beta,\alpha,0},
D_{\beta,\alpha,1},\ldots,D_{\beta,\alpha,N-1}\right]^T$
is just the $N$-point FFT of  vector $\sqrt{N} \bgamma$, where $\bgamma$ is
an $N$-dimensional vector, which is a right cyclic shift of
${\eta}_{max}+M-{\eta}_{\beta,\alpha}-1$ positions of vector
%$\left[d_{\beta,\alpha,0},d_{\beta,\alpha,1},\cdots,
%d_{\beta,\alpha,M-1},\mathbf{0}_{1\times\left({N-M}\right)}\right]^T$,
%i.e.,
$$
\left[d_{\beta,\alpha,0},d_{\beta,\alpha,1},\cdots,
d_{\beta,\alpha,M-1},\underbrace{0,\cdots,0}_{N-M}\right]^T,
$$
where $d_{\beta,\alpha,m}$ are the weighting RCS coefficients,
similar to the single transmitter and single receiver case studied in
\cite{TxzOFDMSAR}.

Then, the pulse compression and coherent integration can be achieved by
performing the $N$-point IFFT operation on  vector
$\hat{\bD}_{\beta,\alpha}=\left[\hat{D}_{\beta,\alpha,0},
\hat{D}_{\beta,\alpha,1},\ldots,\hat{D}_{\beta,\alpha,N-1}\right]^T$
and we obtain:
\begin{equation}\label{tildedrtm}
  \tilde{d}_{\beta,\alpha,\tilde{m}}=
  \frac{1}{\sqrt{N}}\sum_{n=0}^{N-1}
  \hat{D}_{\beta,\alpha,n}\textrm{exp}
  \left\{\frac{j2\pi \tilde{m}n}{N}\right\},\ \tilde{m}=0,1,\ldots, N-1.\\
\end{equation}

So, the estimate of ${d}_{\beta,\alpha,m}$ can be achieved by a left
cyclic shift of ${\eta}_{max}+M-{\eta}_{\beta,\alpha}-1$ positions
of vector $\tilde{d}_{\beta,\alpha,\tilde{m}}$, i.e., vector
$\left[\hat{d}_{\beta,\alpha,0},\ldots,
\hat{d}_{\beta,\alpha,M-1}\right]^T$ is equal to the first $M$
elements of vector
$$
\left[\tilde{d}_{\beta,\alpha,N-{\eta}_{max}-M+{\eta}_{\beta,\alpha}+1},
\ldots,\tilde{d}_{\beta,\alpha,N-1},
\tilde{d}_{\beta,\alpha,0}, \ldots,
\tilde{d}_{\beta,\alpha,N-{\eta}_{max}-M+{\eta}_{\beta,\alpha}}\right]^T.
$$

We then
 obtain the following estimates of the $M$ weighting RCS coefficients at
the $\beta$th receiver due to the $\alpha$th transmitter:
\begin{equation}\label{hatdrtm}
 \hat{d}_{\beta,\alpha,m}=
 {\sqrt{N}}d_{\beta,\alpha,m}+{w}_{\beta,\alpha,m}
 ,\ m=0,1,\ldots, M-1,
\end{equation}
where ${w}_{\beta,\alpha,m}$ is the $m$th output of the $N$-point
IFFT of the vector $\left[\overline{W}_{\beta,\alpha,0},
\overline{W}_{\beta,\alpha,1},\ldots,
\overline{W}_{\beta,\alpha,N-1}\right]^T$ that is the
$\beta$th row and the $\alpha$th column element of matrix
$\overline{\mathbf{W}}_k$ for $k=0,1,\ldots,N-1$.
$\overline{W}_{\beta,\alpha,k}$ can be written as
\begin{equation}\label{overlineWrtk}
  \overline{W}_{\beta,\alpha,k}=\frac{\sum\limits_{p=0}^{P-1}
  W_{\beta,k}^{(p)}S_{\alpha,k}^{(p)}}{\sum\limits_{p=0}^{P-1}
  \left|S_{\alpha,k}^{(p)}\right|^2},\ k=0,1,\ldots,N-1.
\end{equation}

In $(\ref{hatdrtm})$, $d_{\beta,\alpha,m}$ can be recovered without
any interference from other transmitted signals or IRCI from other range
cells. Then, using (\ref{drlm}), we can compensate the phase and obtain the
estimate of the RCS coefficient $g_{\beta,\alpha,m}$ as
\begin{equation}\label{hatgrlm}
\hat{g}_{\beta,\alpha,m}=\hat{d}_{\beta,\alpha,m}\textrm{exp}
\left\{j2\pi
f_c\left[\tau_{\alpha,m}+\tau_{\beta,m}\right]\right\}.\\
\end{equation}

In the above joint pulse compression and coherent integration, the operations
of FFT in (\ref{Urk}), the estimate of $\mathbf{D}_k$ in (\ref{hatbDk}) and
IFFT in (\ref{tildedrtm}) are applied. Thus, we need to analyze the changes of
the noise power in each step of the above range reconstruction method. Assume
that the noise component ${w}_{\beta,n}^{(p)}$ in (\ref{urn}) is a
complex white Gaussian variable with zero-mean and variance $\sigma_n^2$, i.e.,
$w_{\beta,n}^{(p)}\sim\mathcal{CN}\left(0,\sigma_n^2\right)$ for all
receivers $\beta$, pulses $p$ and samples $n$. Since the FFT operation is
unitary, after the process in (\ref{Urk}),  the additive noise power of
$W_{\beta,k}^{(p)}$ does not change, i.e.,
$W_{\beta,k}^{(p)}\sim\mathcal{CN}\left(0,\sigma_n^2\right)$. In the same
way, the noise power of each element in $\mathbf{W}_k$ in (\ref{bUk}) is also
$\sigma_n^2$. However, after the operation for the estimate of $\mathbf{D}_k$
in (\ref{hatbDk}), the variance of a noise component
$\overline{W}_{\beta,\alpha,k}$ in (\ref{overlineWrtk}) can be
calculated as
$$E\left\{\overline{W}_{\beta,\alpha,k}\overline{W}_{\beta,\alpha,k}^{\dagger}\right\}
=\sigma_n^2\left[\sum\limits_{p=0}^{P-1}
\left|S_{\alpha,k}^{(p)}\right|^2\right]^{-1},$$ and thus
$$\overline{W}_{\beta,\alpha,k}\sim\mathcal{CN}
\left(0,\sigma_n^2\left[\sum\limits_{p=0}^{P-1}
\left|S_{\alpha,k}^{(p)}\right|^2\right]^{-1}\right),\ k=0,1,\ldots,N-1,$$
for all $\beta$ and $\alpha$. Moreover, after the IFFT operation in
(\ref{tildedrtm}), we then have finished the joint pulse compression and
coherent integration. The noise power of ${w}_{\beta,\alpha,m}$ in
(\ref{hatdrtm}) is
$$E\left\{{w}_{\beta,\alpha,m}{w}_{\beta,\alpha,m}^{\dagger}\right\}
=\frac{\sigma_n^2}{N}\sum\limits_{k=0}^{N-1}\left[\sum\limits_{p=0}^{P-1}
\left|S_{\alpha,k}^{(p)}\right|^2\right]^{-1}$$ and
$${w}_{\beta,\alpha,m}\sim\mathcal{CN}
\left(0,\frac{\sigma_n^2}{N}\sum\limits_{k=0}^{N-1}\left[\sum\limits_{p=0}^{P-1}
\left|S_{\alpha,k}^{(p)}\right|^2\right]^{-1}\right).$$

Thus, from (\ref{hatdrtm}), we can obtain the SNR of the signal after the joint
pulse compression and coherent integration at the $\beta$th receiver due
to the transmission from the $\alpha$th transmitter and reflected from the
$m$th range cell as,
\begin{equation}\label{SNRrtm}
\textrm{SNR}_{\beta,\alpha,m}=
\frac{N^2\left|d_{\beta,\alpha,m}\right|^2}{\sigma_n^2\sum\limits_{k=0}^{N-1}
\left[\sum\limits_{p=0}^{P-1}
\left|S_{\alpha,k}^{(p)}\right|^2\right]^{-1}}.
\end{equation}
Notice that, a larger $\textrm{SNR}_{\beta,\alpha,m}$ can be
obtained with a smaller value of
$$\sum\limits_{k=0}^{N-1}\left[\sum\limits_{p=0}^{P-1}
\left|S_{\alpha,k}^{(p)}\right|^2\right]^{-1}$$
by designing
$S_{\alpha,k}^{(p)}$. With a given $\alpha$th transmitter and the
energy constraint
$$\sum\limits_{p=0}^{P-1}\sum\limits_{k=0}^{N-1}
\left|S_{\alpha,k}^{(p)}\right|^2=\frac{1}{\mathbb{T}},$$ when
$\sum\limits_{p=0}^{P-1} \left|S_{\alpha,k}^{(p)}\right|^2$ has constant
module for all $k$, i.e.,
\begin{equation}
\sum\limits_{p=0}^{P-1}
\left|S_{\alpha,0}^{(p)}\right|^2=\sum\limits_{p=0}^{P-1}
\left|S_{\alpha,1}^{(p)}\right|^2=\ldots=\sum\limits_{p=0}^{P-1}
\left|S_{\alpha,N-1}^{(p)}\right|^2=\frac{1}{N\mathbb{T}},
\end{equation}
we obtain the minimal value of
$$\sum\limits_{k=0}^{N-1}\left[\sum\limits_{p=0}^{P-1}
\left|S_{\alpha,k}^{(p)}\right|^2\right]^{-1}={N^2\mathbb{T}}.$$
In this
case, the maximal SNR after the joint pulse compression and coherent
integration can be obtained as
\begin{equation}\label{SNRmax}
\textrm{SNR}_{\beta,\alpha,m}^{(max)}=
\max\limits_{\tilde{\emph{\textbf{S}}}_{\alpha}:
\|{\tilde{\emph{\textbf{S}}}_{\alpha}}\|^2=\frac{1}{\mathbb{T}}}
\left\{\textrm{SNR}_{\beta,\alpha,m}\right\}=
\frac{\left|d_{\beta,\alpha,m}\right|^2}{\mathbb{T}\sigma_n^2},
\end{equation}
where
$\tilde{\emph{\textbf{S}}}_{\alpha}=\left[\left(\bS_{\alpha}^{(0)}\right)^T,\ldots,\left(\bS_{\alpha}^{(P-1)}\right)^T\right]^T
\in\mathbb{C}^{{PN}\times{1}}$.

Thus, for the $\alpha$th transmitter, the optimal signal
$S_{\alpha,k}^{(p)}$ should satisfy a requirement that the transmitted
energy summations of the $P$ pulses within a CPI, i.e.,
$\sum\limits_{p=0}^{P-1} \left|S_{\alpha,k}^{(p)}\right|^2$, have constant
module for all $k$. Otherwise, the SNR after the range reconstruction will be
degraded. Here, we define the SNR degradation factor as
\begin{equation}\label{xi}
\xi=\frac{\textrm{SNR}_{\beta,\alpha,m}}
{\textrm{SNR}_{\beta,\alpha,m}^{(max)}}
=\frac{N^2\mathbb{T}}{\sum\limits_{k=0}^{N-1} \left[\sum\limits_{p=0}^{P-1}
\left|S_{\alpha,k}^{(p)}\right|^2\right]^{-1}}.
\end{equation}
Notice that $\xi\in\left(0,1\right]$ is independent of the noise power
$\sigma_n^2$ and the weighting RCS coefficient $d_{\beta,\alpha,m}$. Since we assume that the row
vectors of matrix $\mathbf{S}_k$ are orthogonal each other and have the same
norm, the above degradation factor $\xi$ is also independent of $\beta$
and $\alpha$. The SNR degradation factor
$\xi$ in (\ref{xi}) is for the  performance
of both  pulse compression and coherent integration of all the $P$ pulses within a
CPI, but, not only the pulse compression of a single pulse in
\cite{TxzAPLOFDMSAR}.

We recall that the number of the OFDM signal subcarriers should satisfy
$N\geq{\eta}_{max}+M$. Thus, the length of the transmitted signals should be
increased with the increases of the width $R_w$ for the radar footprints in the
range direction and/or ${\eta}_{max}$. The pulse length will be much longer than the traditional
radar pulse for a wide width $R_w$ (or large $M$)
and/or a large delay ${\eta}_{max}$, which may be a problem,
especially, for covert/military radar applications. Meanwhile, the CP removal
for the elimination of the interference at the receivers may cause high
transmitted energy redundancy as we have mentioned in \cite{TxzAPLOFDMSAR}.
Therefore, it is necessary for us to achieve MIMO OFDM radar with arbitrary
pulse length that is independent of $R_w$. The main idea is to generate $P$
pulses $s_\alpha^{(p)}(t),\ t\in \left[pT_r,pT_r+T+T_{GI}\right],\
p=0,1,\ldots,P-1$, for all $\mathbb{T}$ transmitters, such that the discrete
time sequence of $s_\alpha^{(p)}(t),\ pT_r\leq t\leq pT_r+T+T_{GI}$:
$s_{\alpha,i}^{(p)}=s_{\alpha}^{(p)}(iT_s),\ 0\leq i\leq
N+{\eta}_{max}+M-2$ in (\ref{sti}), is zero at the head and the tail parts as
\begin{equation}\label{headtailends}
\left[s_{\alpha,0}^{(p)}, \ldots,s_{\alpha,{\eta}_{max}+M-2}^{(p)}\right]^T=
\left[s_{\alpha,N}^{(p)}, \ldots,s_{\alpha,N+{\eta}_{max}+M-2}^{(p)}\right]^T=
\mathbf{0}_{\left({\eta}_{max}+M-1\right)\times 1}.
\end{equation}
In the meantime, $s_{\alpha,i}^{(p)}$ is also a sampled discrete time
sequence of an OFDM pulse in (\ref{OFDM}) for $t\in
\left[pT_r,pT_r+T+T_{GI}\right]$.
This zero head and tail condition (\ref{headtailends}) is the same as that
in \cite{TxzAPLOFDMSAR}.
Then, in this case, the continuous time signal
$s_{\alpha}^{(p)}(t)$ is only transmitted
on the time interval $t\in \left[pT_r+T_{GI},pT_r+T\right]$ that
has length $T-T_{GI}$, where $T_{GI}$ is the length of the guard interval
and also the zero-valued
head part  of the signal that  leads to the zero-valued CP part at the tail.
Since $T_{GI}$ can be arbitrarily designed, the OFDM pulse length
$T-T_{GI}$ can be arbitrary as well.
For more details, we refer to \cite{TxzAPLOFDMSAR}.
Based on the above analysis, the key task of
the following section is the design of these multiple OFDM sequences.

\section{Design of Multiple OFDM Sequences}\label{OFDM_Design}
In this section, we design the weight sequences in  the $P$  OFDM pulses for each
transmitter, i.e., the matrix $\mathbf{S}_k=[S_{\alpha,k}^{(p)}]_{1\leq \alpha
\leq \alpha, 0\leq p\leq P-1}$ for $k=0,1,..., N-1$ in (\ref{bfSk}).
There are three indices here: one is the transmitter index $\alpha$, one is
the OFDM pulse index $p$ for each transmitter, and the third one is the subcarrier index $k$.
We start with the design criterion.

\subsection{Design criterion}
Any segment of an OFDM pulse in (\ref{OFDM}) is determined by a weight sequence
$\bS_{\alpha}^{(p)}=\left[S_{\alpha,0}^{(p)},
S_{\alpha,1}^{(p)},\right.$
$\ldots,$ $\left. S_{\alpha,N-1}^{(p)}\right]^T$ that is
determined by its $N$-point IFFT
${\bs_{\alpha}^{(p)}}=\left[s_{\alpha,0}^{(p)},
s_{\alpha,1}^{(p)},\ldots,s_{\alpha,N-1}^{(p)}\right]^T$. Thus, the
design of ${\bs_{\alpha}^{(p)}}$ is equivalent to the design of
$\bS_{\alpha}^{(p)}$. Based on the above discussions,   ${\bs_{\alpha}^{(p)}}$ and
$\bS_{\alpha}^{(p)}$ should satisfy the following conditions:
\begin{itemize}
\item[1)]
{\bf Frequency domain orthogonality among transmitters for every subcarrier}.
 As it was mentioned earlier, in order not to
enhance the noise in the estimate in (\ref{hatbDk}) for RCS coefficients,
matrix $\mathbf{S}_k$ has to be a flat unitary matrix, i.e.,
$\mathbf{S}_k\mathbf{S}_k^{\dagger}=\mathbf{I}_{\mathbb{T}}$ for each
$k=0,1,\ldots,N-1$. Specifically, the sequence $\bS_{\alpha,k}$ should be
orthogonal to sequence $\bS_{\tilde{\alpha},k}$ for different transmitters
$\alpha\neq \tilde{\alpha}$ and
$1\leq\alpha,\tilde{\alpha}\leq \mathbb{T}$, and have the
same norm, where
$\bS_{\alpha,k}=\left[S_{\alpha,k}^{(0)},S_{\alpha,k}^{(1)},
\ldots,S_{\alpha,k}^{(P-1)}\right]$ is the $\alpha$th row of
$\mathbf{S}_k$. Note that this orthogonality is for every subcarrier
in the discrete frequency
domain of the signal waveforms but not in the time domain as commonly used in a
MIMO radar. The advantage of this  orthogonality in the frequency domain is
that it is
not affected by time delays in the time domain, while the orthogonality in the
time domain is sensitive to any time delays. In addition, this discrete
orthogonality in the frequency domain does not require that the frequency bands
of the waveforms do not overlap each other as commonly used in the frequency
division MIMO radar and in fact, all the frequency bands of the $\mathbb{T}$
waveforms can be the same. It implies that the range resolution is not
sacrificed as what is done in frequency division MIMO radar. This criterion
deals with the transmitter index $\alpha$ and the OFDM pulse index $p$,
and the subcarrier index $k$ is free.
\item[2)]
{\bf Zero head and tail condition}. Sequence ${\bs_{\alpha}^{(p)}}$ should
satisfy the zero head and tail condition in (\ref{headtailends}) for all $p$
and $\alpha$. This criterion only deals with the time index $i$ in a
pulse, or equivalently, the subcarrier index $k$.
\item[3)]
{\bf Flat total spectral power of $P$ pulses}. To avoid the SNR degradation as
the estimation of the weighting
RCS coefficients in (\ref{hatbDk}) and what follows, and achieve the maximal SNR after
pulse compression and coherent integration, for the $\alpha$th
transmitter, the transmitted energy summation of all the $P$ pulses within a
CPI should have constant module for all $k$, i.e.,
$$\sum\limits_{p=0}^{P-1}
\left|S_{\alpha,k}^{(p)}\right|^2=\frac{1}{N\mathbb{T}}.$$
This criterion only deals with the pulse index $p$.
\item[4)]
{\bf Good PAPR property}. The PAPR of the transmitted OFDM pulse
$s_\alpha^{(p)}(t),\ p=0,1,\ldots,P-1$, in (\ref{OFDM}) for $t\in
\left[pT_r+T_{GI},pT_r+T\right]$ should be minimized for an easy practical
implementation of the radar. This criterion also only deals with the time index
$t$ in a pulse, or equivalently, the subcarrier index $k$.

\end{itemize}

The basic idea of the following designs to satisfy the above four criteria
is to first use a pattern (called orthogonal design)
 of placing $P$  pulses to ensure the
orthogonality condition 1) among all the $\mathbb{T}$ transmitters, where the $P$  pulses and/or their
complex conjugates and/or their shifted versions
etc.  are used by every transmitter. After this is done, it is only needed to
work on these $P$ pulses to satisfy the other three criteria above, which are independent of a transmitter.

\subsection{Frequency domain orthogonality using orthogonal designs}

The orthogonality condition 1) for the weighting matrix
$\mathbf{S}_k$ in (\ref{bfSk}) is for all subcarrier indices $k$,
i.e., it is for a matrix whose entries are variables but not simply
constants. This motivates us to use
complex orthogonal designs (COD)
\cite{AlamoutiASimpleIEEEJSAC19981451,TarokhSpaceTimeBlockIEEETIT19991456,
SuWeiSpaceTimeWPC20031, XueBinOnTheNonexistenceIEEETIT20032984,
HaiQuanUpperBoundsIEEETIT20032788,XueBinOrthogonalDesignsIEEETIT20032468,
SuWeiASystematicDesignIEEECL2004380,KejieClosedFormDesignsIEEETIT20054340} whose entries are
arbitrary complex variables. Furthermore, each row vector of a COD
uses the same set of complex variables, which corresponds to that each
 transmitter
uses the same set of OFDM pulses and therefore we only need to consider
$P$  pulses for one transmitter as explained above.

Let us briefly recall a COD
\cite{AlamoutiASimpleIEEEJSAC19981451,TarokhSpaceTimeBlockIEEETIT19991456,
SuWeiSpaceTimeWPC20031, XueBinOnTheNonexistenceIEEETIT20032984,
HaiQuanUpperBoundsIEEETIT20032788,XueBinOrthogonalDesignsIEEETIT20032468,
SuWeiASystematicDesignIEEECL2004380,KejieClosedFormDesignsIEEETIT20054340}. A
$\mathbb{T} \times P$ COD\footnote{The COD definition we use in this paper
follows the original COD definition
\cite{TarokhSpaceTimeBlockIEEETIT19991456,XueBinOrthogonalDesignsIEEETIT20032468}
where no linear combinations or repetitions of  complex variables $x_i$ is
allowed in the matrix entry or any row of the matrix. This appears important in
the applications in this paper. More general COD definitions can be found in
\cite{TarokhSpaceTimeBlockIEEETIT19991456,XueBinOrthogonalDesignsIEEETIT20032468,
HaiQuanUpperBoundsIEEETIT20032788,KejieClosedFormDesignsIEEETIT20054340} where
any complex linear combinations of complex variables $x_i$ are allowed in the
entries of the matrix and does not affect their applications in wireless MIMO
communications.} with $P_0$ complex variables $x_1, x_2,..., x_{P_0}$ is a
$\mathbb{T} \times P$ matrix $\mathbf{X}$ such that its every entry is either
$0$, $x_i$, $-x_i$, $x_i^*$, or $-x_i^*$ and satisfies the following identity
\begin{equation}\label{cod}
\mathbf{X}\mathbf{X}^{\dagger}= (|x_1|^2+\cdots +|x_{P_0}|^2)
 \mathbf{I}_{\mathbb{T}},
\end{equation}
where every $x_i$ may take any complex value. CODs have been used for
orthogonal space-time block codes (OSTBC) in MIMO communications to collect
full spatial diversity with fast maximum-likelihood (ML) decoding, see for
example
\cite{AlamoutiASimpleIEEEJSAC19981451,TarokhSpaceTimeBlockIEEETIT19991456,
SuWeiSpaceTimeWPC20031, XueBinOnTheNonexistenceIEEETIT20032984,
HaiQuanUpperBoundsIEEETIT20032788,XueBinOrthogonalDesignsIEEETIT20032468,
SuWeiASystematicDesignIEEECL2004380,KejieClosedFormDesignsIEEETIT20054340}.
Note that, as we shall see later, our use of a COD in the following is not from
an OSTBC point of view but only from the structured orthogonality (\ref{cod}).
A closed-form inductive design of a $\mathbb{T} \times P$ COD for any
$\mathbb{T}$ is given in \cite{KejieClosedFormDesignsIEEETIT20054340}. The
following are two simplest but non-trivial COD for $\mathbb{T}=2$ and $4$,
respectively,
\begin{equation}\label{cod24}
\mathbf{X}_2=\left[ \begin{array}{cc}
x_1 & x_2\\
-x_2^* & x_1^*
\end{array}
\right]
\,\,\, \mbox{ and }\,\,\,
\mathbf{X}_4=\left[ \begin{array}{cccc}
x_1 & x_2 & x_3 & 0\\
-x_2^* & x_1^* & 0 & x_3\\
-x_3^* & 0 & x_1^* & -x_2\\
 0 & -x_3^* & x_2^* & x_1
\end{array}
\right].
\end{equation}
The above COD $\mathbf{X}_2$ was first used as an OSTBC by Alamouti
in \cite{AlamoutiASimpleIEEEJSAC19981451} and it is now well-known as Alamouti code in MIMO communications.
From the second example $\mathbf{X}_4$ above, one may see that
the number, $P_0$, of the nonzero variables in a COD may not
be necessarily equal
to the number, $P$,  of its columns.
In fact, for a given $\mathbb{T}$, the relationship between
$P$, $P_0$ and $\mathbb{T}$ has been given in \cite{XueBinOrthogonalDesignsIEEETIT20032468,KejieClosedFormDesignsIEEETIT20054340}, where it is shown that
\begin{equation}\label{codrate}
\frac{P_0}{P}=\frac{\lceil \frac{\mathbb{T}}{2} \rceil +1}{2 \lceil \frac{\mathbb{T}}{2} \rceil}
\end{equation}
is achieved with closed-form designs  in
\cite{KejieClosedFormDesignsIEEETIT20054340}.
From the COD definition, it is not hard to see that
every row of a COD contains the same set of compex variables
$x_1$,..., $x_{P_0}$ and every such a variable $x_i$ only appears once.
With this property, when we apply a COD as a weighting matrix $\mathbf{S}_k$
for every $k$,
among the $P$ pulses, only $P_0$ non-zero OFDM pulses are used for every
transmitter and the other $P-P_0$
pulses are all zero-valued.

With a COD, we may design a weighting matrix $\mathbf{S}_k$ for every $k$.
Let us use the above $2\times 2$ COD as an example. It is used for the
case of $\mathbb{T}=P=2$.
The corresponding $2\times 2$ weighting matrix $\mathbf{S}_k$ for every $k$ is
\begin{equation}\label{Alamouti}
\left[ \begin{array}{c}
\bS_{1,k}^T\\
\bS_{2,k}^T
\end{array}\right]=
\begin{bmatrix}
S_{1,k}^{(0)} &S_{1,k}^{(1)}\\
S_{2,k}^{(0)} &S_{2,k}^{(1)}
\end{bmatrix}
=
\begin{bmatrix}
 S_{1,k}^{(0)} &S_{1,k}^{(1)}\\
-\left(S_{1,k}^{(1)}\right)^*
&\left(S_{1,k}^{(0)}\right)^*
\end{bmatrix},~k=0,1,\ldots,N-1.
\end{equation}
Then, $\bS_{1,k}$ and $\bS_{2,k}$ are orthogonal
and have the same norm for every $k$.
% The
%orthogonal code structures for $\mathbb{T}>2$ and $P>2$ have been investigated
%in \cite{TarokhSpaceTimeBlockIEEETIT19991456,SuWeiSpaceTimeWPC20031,
%XueBinOnTheNonexistenceIEEETIT20032984,
%HaiQuanUpperBoundsIEEETIT20032788,XueBinOrthogonalDesignsIEEETIT20032468,
%SuWeiASystematicDesignIEEECL2004380,KejieClosedFormDesignsIEEETIT20054340}.
The discrete time domain sequences ${\bs_{\alpha}^{(p)}}=
\left[s_{\alpha,0}^{(p)},\ldots,s_{\alpha,N-1}^{(p)}\right]^T$
for the $\alpha$th transmitter and the $p$th OFDM pulse is
%and ${\bs_{\tilde{\alpha}}^{(p)}}=\left[s_{\tilde{\alpha},0}^{(p)},\ldots,s_{\tilde{\alpha},N-1}^{(p)}\right]^T$
 obtained by taking  the $N$-point IFFT of ${\bS_{\alpha}^{(p)}}=[S_{\alpha,0}^{(p)},\ldots, S_{\alpha,N-1}^{(p)}]^T$.
From the above design in (\ref{Alamouti}) for two transmitters,
the two OFDM pulses for the first transmitter are free to design so far,
while the two OFDM pulses for the second  transmitter
in the frequency domain are determined by the two pulses
for the first transmitter.
The two OFDM pulses for the second transmitter in the discrete time domain
are, correspondingly,
$$
s_{2,i}^{(0)}= - \left(s_{1,N-i}^{(1)}\right)^*
\,\,\,\mbox{ and }\,\,\,
s_{2,i}^{(1)}=\left(s_{1,N-i}^{(0)}\right)^*,\,\,\,
i=0,1,...,N-1.
$$
In the continuous time domain, they are
$$
s_{2}^{(0)}(t)= - \left(s_{1}^{(1)}(T-t)\right)^*
\,\,\,\mbox{ and }\,\,\,
s_{2}^{(1)}(t)=\left(s_{1}^{(0)}(T-t)\right)^*,
$$
where $t\in [T_{GI}, T+T_{GI}]$ when the CP is not included and
$t\in [0, T+T_{GI}]$ when the CP is included.

For  general $\mathbb{T}$  transmitters, from a COD design
\cite{KejieClosedFormDesignsIEEETIT20054340}, such as (\ref{cod24}) for
$\mathbb{T}=4$, the discrete complex weight sequences for the first transmitter
${\bS_{1}^{(p)}} = [S_{1,0}^{(p)},\ldots , S_{1, N-1}^{(p)}]^T$ are either the
all zero sequence ($P-P_0$ of them) or free to design ($P_0$ of them) so far
(more conditions will be imposed for the other criteria 2)-4) later). The
discrete complex weight sequences for any other transmitter $\bS_{\alpha}^{(p)}
= [S_{\alpha,0}^{(p)},\ldots, S_{\alpha, N-1}^{(p)}]^T$  for $\alpha>1$ are
either the all zero sequence ($P-P_0$ of them as the first transmitter), or
$\pm {\bS_{1}^{(p')}}$, or $\pm \left( {\bS_{1}^{(p')}}\right)^*$ for some $p'$
with $0\leq p'\neq p\leq P-1$. Then, the discrete time domain sequences for any
other transmitter $s_{\alpha,i}^{(p)}$ for $\alpha>1$ are either the all zero
sequence or $\pm [s_{1,i}^{(p')}]_{0\leq i\leq N-1}$ or $\pm \left(
[s_{1,N-i}^{(p')}]_{0\leq i\leq N-1} \right)^*$ for some $p'$ with $0\leq
p'\neq p\leq P-1$. In the continuous time domain, a pulse transmitted by any
other transmitter $s_{\alpha}^{(p)}(t)$ for $\alpha>1$ are either the all
zero-valued pulse, or  $\pm s_{1}^{(p')}(t)$ or $\pm \left( s_{1}^{(p')}(T-t)
\right)^*$ for some $p'$ with $0\leq p'\neq p\leq P-1$. Note that for the
notational convenience, all the above $P$ pulses are considered over the same
time interval. However, these $P$ pulses are arranged in sequential in time
after they are designed and when they are used/transmitted.

In the case of $\mathbb{T}=4$ in (\ref{cod24}), $P_0=3$ and $P=4$ and there is
one all zero pulse for each transmitter and at any time, only three
transmitters transmit signals and  the idle transmitter alternates.

From the above pulse placement among transmitters using a COD, the transmitted
pulses for the first transmitter are either all zero-valued, or free to design,
and the pulses transmitted by any other transmitters are the pulses transmitted
by the first transmitter possibly with some simple operations of negative
signed, complex conjugated, and/or time-reversed in the pulse period, and no
more and no less pulses are transmitted. These operations do not change the
signal power in frequency domain or the signal PAPR in time domain for a pulse,
and thus do not change the conditions 3) and 4) of the design criteria studied
above. So, for the design criteria 3) and 4), we only need to consider the
$P_0$ non-zero pulses for the first transmitter. Note that the complex
conjugation in frequency domain not only causes the complex conjugation in time
domain but also causes the time reversal in time domain as expressed above. The
time reversal operation to a pulse in time domain may change the zero head and
tail condition 2) in the above design criteria, i.e., if a sequence satisfies
the zero head and tail condition (\ref{headtailends}), its time-reversed
version   may not satisfy the zero head and tail condition (\ref{headtailends})
anymore. However, if  sequence $\bs_{\alpha}^{(p)}$, with its FFT
$\bS_{\alpha}^{(p)}$, satisfies not only the condition in (\ref{headtailends})
but also
\begin{equation}\label{conjugateCond}
\left[s_{\alpha,N-{\eta}_{max}-M+2}^{(p)},
\ldots,s_{\alpha,N-1}^{(p)}\right]^T=
\mathbf{0}_{\left({\eta}_{max}+M-2\right)\times 1},
\end{equation}
then, not only sequence $\bs_{\alpha}^{(p)}=[s_{\alpha, i}^{(p)}]$ satisfies
the zero head and tail condition (\ref{headtailends}) but also its time
reversed version $[s_{\alpha, N-i}^{(p)}]$ also satisfies  the zero head and
tail condition (\ref{headtailends}). Due to this additional zero-segment
condition in (\ref{conjugateCond}), the PAPR in time domain should be
re-defined as the PAPR only over the non-zero portion, i.e., the portion for
$t\in \left[pT_r+T_{GI},\right.$ $\left.pT_r+T-T_{GI}+T_s\right]$, of a pulse.
Therefore, the design criteria  2) and 4) should be updated as:
\begin{itemize}
\item [2)] {\bf New zero head and tail condition}.
Sequence ${\bs_{\alpha}^{(p)}}$ should satisfy the
zero head and tail  conditions in (\ref{headtailends}) and
(\ref{conjugateCond}) for all $p$ and $\alpha$.
\item [4)] {\bf New good PAPR property}.
The PAPR of the transmitted non-zero-valued OFDM pulse $s_\alpha^{(p)}(t)$ for each $p$,
$p=0,1,\ldots,P_0-1$,
and each $\alpha$, $1\leq \alpha\leq \mathbb{T}$,
in (\ref{OFDM}) for $t\in
\left[pT_r+T_{GI},\right.$ $\left.pT_r+T-T_{GI}+T_s\right]$ should be minimized.
\end{itemize}

In this case, with the conditions
in (\ref{headtailends}) and
(\ref{conjugateCond}), a transmitted time domain sequence of the
$\alpha$th transmitter and the $p$th pulse becomes
$\tilde{\bs}_{\alpha}^{(p)}=\left[s_{\alpha,{\eta}_{max}+M-1}^{(p)},
s_{\alpha,{\eta}_{max}+M}^{(p)},
\ldots,s_{\alpha,N-{\eta}_{max}-M+1}^{(p)}\right]^T\in
\mathbb{C}^{N_t\times 1}$ for $1\leq\alpha\leq \mathbb{T}$ and
$0\leq{p}\leq{P-1}$, where $N_t=N-2{\eta}_{max}-2M+3$ is the length of the
transmitted non-zero OFDM sequences.
Among these $P$ pulses, only $P_0$ of them are not all zero pulses.
Thus, the normalized transmitted energy constraint
of $\tilde{\bs}_{\alpha}^{(p)}$ is that the mean transmitted power of
$\tilde{\bs}_{\alpha}^{(p)}$ is $\frac{1}{N_t{\mathbb{T}}P_0}$.
Hence, the
SNR of the received signal from the $m$th range cell before pulse compression
and coherent integration is
\begin{equation}\label{OLSNRrtm}
\overline{\textrm{SNR}}_{\beta,\alpha,m}=
\frac{\left|d_{\beta,\alpha,m}\right|^2}{N_t{\mathbb{T}}P_0\sigma_n^2}.
\end{equation}
Note that the maximal SNR of the $m$th range cell after the joint pulse
compression and coherent integration
$\textrm{SNR}_{\beta,\alpha,m}^{(max)}$ in (\ref{SNRmax}) is equal
to $P_0N_t\overline{\textrm{SNR}}_{\beta,\alpha,m}$, and the SNR gains
of the pulse coherent integration $P_0$ (the number of
non-zero pulses)  and the pulse compression $N_t$ (the non-zero-valued pulse
length) are
consistent with the traditional radar applications
\cite{skolnik2001Introduction}. Based on the above analysis, the key task of
the remainder of this section is to design a sequence $\bs_{\alpha}^{(p)}$
that simultaneously satisfies the above criteria 2), 3) and 4).

Before finishing this subsection, a remark on using a COD in the above pulse
placement among transmitters is follows. When the number $\mathbb{T}$
of transmitters is not small, either the number $P$ of pulses will be
much larger than $\mathbb{T}$
or the number $P_0$ of non-zero pulses can be put in will be small.
There is a tradeoff among these three parameters as we have mentioned
earlier for a COD design.
When $P_0$ is small, there are less degrees
 of freedom in the pulse design, which
will affect the MIMO OFDM radar performance, when other conditions are imposed
as we shall see later. Furthermore, when $P_0$ is small, the radar transmitter
usage is low and may not be preferred in radar applications. From the COD rate
property (\ref{codrate}), one can see that $P_0$ is always more than $P/2$,
i.e., among a CPI of  $P$ pulses, there are always more than half of $P$ pulses
are non-zero OFDM pulses. A trivial unitary matrix $\mathbf{S}_k$ in
(\ref{bfSk}) is a diagonal matrix with all diagonal elements of the same norm.
This corresponds to the case when there is only one transmitter transmits at
any time in a CPI and then the radar transmitter usage becomes the lowest,
which is again not preferred. On the other hand, when $P$ is large, the time to
transmit these $P$ pulses becomes long, which may not be preferred in some
radar applications either. Another remark is that unitary  matrices  ${\mathbf
S}_k$ have been also constructed in \cite{WuMIMOOFDMIETRSN201028} where all
unitary matrices ${\mathbf S}_k$ for all $k$ are from a single constant unitray
matrix and each ${\mathbf S}_k$ for each $k$ has only one free parameter on
phase. This may limit the ability to find desired waveforms with some
additional desired properties, such as those we will discuss next.

Also in what follows, for the notational convenience,
we use $P$ instead of $P_0$ to denote the number of non-zero
OFDM pulses to design since an all-zero-valued pulse does not affect the other
pulses.

\subsection{Flat total spectral power
 using paraunitary filterbanks}\label{paraunitaryfilterbank}

From the above studies, we only need to design $P$ pulses for the first
transmitter. In this subsection, we design $P$ OFDM pulses by designing their
equivalent OFDM sequences $\bs^{(p)}$ in time domain or $\bS^{(p)}$ in
frequency domain, for $p=0,1,\ldots,P-1$, that satisfy the design criteria 2)
(new)  and 3) precisely. We omit their transmitter index $1$ for
convenience. The main idea is to apply the paraunitary filterbank theory
\cite{PPVaidyanathanMultirate1993}
($\!\!$\cite{XiangGenXiaMultirateFilterbanks} for a short tutorial) as follows.

Considering the above criterion 2) (new), the complex weight sequences
${\bS^{(p)}}$, for $p=0,1,\ldots,P-1$, can be written as
\begin{equation}\begin{split}\label{Stkp}
S_{k}^{(p)}=\frac{1}{\sqrt{N}}\sum_{i=\eta_{1\mathrm{st}}}^{N-\eta_{1\mathrm{st}}}
s_{i}^{(p)}\textrm{exp}\left\{-\frac{j2\pi ik}{N}\right\},
~k=0,1,\ldots,N-1,
\end{split}\end{equation}
where $\eta_{1\mathrm{st}}={\eta}_{max}+M-1$ is the index of the first non-zero
value of sequence $\bs^{(p)}$.
Then,  we have $S_{k}^{(p)}=\left. S^{(p)}(z) \right|_{z=W_k}$ for $k=0,1,...,N-1$, where $W_k \stackrel{\Delta}{=}
\textrm{exp}\left\{\frac{j2\pi k}{N}\right\}$ and
\begin{equation}\label{Stzp1}
S^{(p)}(z) =
\frac{z^{-\eta_{1\mathrm{st}}}}{\sqrt{N}}\sum_{i=0}^{N_t-1}
s_{\eta_{1\mathrm{st}}+i}^{(p)} z^{-i},
\end{equation}
where we recall that $N_t=N-2{\eta}_{max}-2M+3$ is the length of the
transmitted non-zero OFDM sequences.
Then, the flat total spectral power in the criterion 3) can be re-written as
\begin{equation}\label{flatpower_dis}
\left. \sum_{p=0}^{P-1} |S^{(p)}(z)|^2  \right|_{z=W_k}= \frac{1}{N\mathbb{T}},
\,\,\,k=0,1,..., N-1.
\end{equation}
The above identity for all  $k$ is ensured by
the following identity on the whole unit circle of $z$,
\begin{equation}\label{flatpower_z}
\sum_{p=0}^{P-1} |S^{(p)}(z)|^2 = \frac{1}{N\mathbb{T}},  \,\,\, |z|=1.
\end{equation}
This identity tells us that if $S^{(p)}(z)$, $p=0,1,...,P-1$,
form a filterbank, then this filterbank can be systematically
constructed by a paraunitary filternbank
with polyphase representations of $P$ filters  $S^{(p)}(z)$, $p=0,1,...,P-1$,
\cite{PPVaidyanathanMultirate1993}
 as follows.
For each $p$, re-write $S^{(p)}(z)$ as
\begin{equation}\label{Stzp2}
S^{(p)}(z)=z^{-\eta_{1\mathrm{st}}} \sum_{q=0}^{P-1}
z^{-q}S_q^{(p)}(z^{P}),
\end{equation}
where
\begin{equation}\label{Stpzkp}
S_{q}^{(p)}(z)=\frac{1}{\sqrt{N}}\sum_{i=0}^{\lceil \frac{N_t-P}{P}\rceil}
s_{\eta_{1\mathrm{st}}+Pi+q}^{(p)}z^{-i}
\end{equation}
is the $q$th polyphase component of $S^{(p)}(z)$.
Clearly, a filter $S^{(p)}(z)$ and its $P$ polyphase components
$S_q^{(p)}(z)$, $q=0,1,...,P-1$,
can be equivalently and easily  converted to each other as above.
These $P^2$ polyphase
components for all the $P$ filters form a $P\times P$ polyphase matrix
$\mathbf{S}(z)=[S_q^{(p)}(z)]_{0\leq p\leq P-1, 0\leq q\leq P-1}$.
Then, the flat spectral power condition (\ref{flatpower_z})
is equivalent to the  losslessness (or paraunitariness)
of the $P\times P$ matrix
$\mathbf{S}(z)\tilde{\mathbf{S}}(z)= \frac{1}{N\mathbb{T}} \mathbf{I}_P$
for all complex values $|z|=1$ (or all complex values $z$ and then this matrix
is called a paraunitary matrix)
\cite{PPVaidyanathanMultirate1993}, where
$\tilde{\mathbf{S}}(z)$ is the tilde operation of $\mathbf{S}(z)$, i.e.,
$\tilde{\mathbf{S}}(z)=\mathbf{S}^{\dagger}(z^{-1})$.
Such a paraunitary matrix can be factorized as
\cite{PPVaidyanathanMultirate1993}:
\begin{equation}\label{FrequenceMatrix}
\mathbf{S}(z)
=
\frac{1}{\sqrt{N\mathbb{T}}}\prod\limits_{l=1}^{\lceil
\frac{N_t-P}{P}\rceil}\bV_l (z) \bV,
\end{equation}
where $\bV$ is a $P\times P$ constant
unitary matrix and
\begin{equation}\label{factor}
\bV_l(z)=\mathbf{I}_P-\bv_l\bv_l^{\dagger}+z^{-1}\bv_l\bv_l^{\dagger},
\end{equation}
where
$\bv_l\in\mathbb{C}^{P\times {1}}$ is a $P$ by $1$ constant
column vector of unit norm.

In order to construct OFDM sequences $\bs^{(p)}$ that satisfy
the new zero head and tail condition 2),
 when
$\frac{N_t-P}{P}$ is not an integer, the above paraunitary
matrix $\mathbf{S}(z)$ can be constructed as
\begin{equation}\label{construction}
\mathbf{S}(z)
=
\frac{1}{\sqrt{N\mathbb{T}}}\prod\limits_{l=1}^{\lfloor
\frac{N_t-P}{P}\rfloor}\bV_l (z) \bV,
\end{equation}
where $\bV$ and  $\bV_l(z)$ are as in (\ref{FrequenceMatrix}) and (\ref{factor}),
respectively. After a paraunitary matrix $\mathbf{S}(z)=[S_q^{(p)}(z)]$ is constructed in (\ref{construction}), we can form $S^{(p)}(z)$ for $p=0,1,...,P-1$ via (\ref{Stzp2}).
Then, sequences $S_k^{(p)}$, $k=0,1,...,N-1$, for $p=0,1,...,P-1$,
satisfy the flat total
spectral power condition 3). The  discrete time domain OFDM sequences
$\bs^{(p)}$ can be obtained by taking the $N$-point IFFT of
$\bS^{(p)}$ for every $p=0,1,...,P-1$, which satisfy the new zero head and tail condition 2).
In this construction, there are $P^2$ complex-valued parameters
in the unitary matrix $\bV$ and $\lfloor \frac{N_t-P}{P}\rfloor \times P$ complex-valued parameters in the $P\times 1$ vectors $\bv_l$ with unit norm for $l=1,2,...$, $\lfloor \frac{N_t-P}{P}\rfloor$. Therefore, there are total
$$
P^2+\lfloor \frac{N_t-P}{P}\rfloor \times P
\approx N_t +P^2 -P
$$
complex-valued parameters to choose  under the constraints
of $\bV \bV^{\dagger}=\mathbf{I}_P$ and $\|\bv_l\|=1$.
As a remark, compared to the single OFDM pulse case studied for single
transmitter radar in \cite{TxzOFDMSAR,TxzAPLOFDMSAR},
i.e., $P=1$, the flat total spectral
power 4) for $P>1$ is easier to achieve.

In order to design OFDM pulses to satisfy the criterion 4), i.e., to
have low PAPR in the time domain, unfortunately, there is no closed-form
construction
(see, for example, a tutorial
  \cite{SeungIEEEWCPAPRoverview200556} for PAPR issues)
as for the previous three criteria 1)-3). One way to design good PAPR pulses
satisfying 1)-3) is to search the above parameters in $\bV$ and $\bv_l$.
However, since there are too many complex-valued parameters to search, it is
hard to find OFDM pulses that satisfy 1)-3) and have good PAPR property in time
domain. Let us go back to re-exam the flat total spectral power property 3)
that is used to achieve the optimal SNR after the joint pulse compression and
coherent integration as what is studied in (\ref{SNRrtm})-(\ref{SNRmax}). In
practice, a small SNR degradation with $\xi\approx 1$ in (\ref{xi}) may not
impact the radar performance much by slightly relaxing the flat total spectral
power condition 3). With this small relaxation, i.e.,
$\sum\limits_{p=0}^{P-1}\left|S_{k}^{(p)}\right|^2
\approx\frac{1}{N\mathbb{T}}$ for all $k=0,1,...,N-1$, it will be much easier
to achieve good PAPR criterion 4) as we shall see below.

\subsection{OFDM sequence design using MICF}\label{OFDM_Design_MICF}
A simple method was proposed in \cite{TxzAPLOFDMSAR} for single OFDM pulse
design, in which the filtering and clipping operations were iteratively applied
in  time and frequency  domains to reduce the PAPR of the transmitted OFDM
pulse and make the complex weights of different subcarriers to be as constant
as possible. Since the above requirements 2), 3) and 4) are respectively
similar\footnote{The difference is that an additional condition of
(\ref{conjugateCond}) is added in the above requirement 3) of this paper.} to
the corresponding requirements 1), 2) and 3) in \cite{TxzAPLOFDMSAR}, by using
the method in \cite{TxzAPLOFDMSAR}, a simple method to achieve
$\sum\limits_{p=0}^{P-1}\left|S_{k}^{(p)}\right|^2
\approx\frac{1}{N\mathbb{T}}$ and the zero head and tail condition 2) is to
design each individual sequence $S_{k}^{(p)}$ for each $p$ separately for
 approximately
constant module $S_{k}^{(p)}$ for all $k$ and $p$, i.e.,
$\left|S_{k}^{(p)}\right|\approx{\frac{1}{\sqrt{N\mathbb{T}P}}}$. However, with
this simple method, there are less degrees of freedom than that when all $P$
pulses are jointly considered in the design, which can be evidenced by
observing that there are closed-form solutions to achieve the flat total
spectral power when $P>1$ as what is studied in the preceding subsection, while
it is much harder (if not impossible) when $P=1$. In the meantime, there are
more degrees of freedom for filtering and clipping when all  $P$ OFDM pulses
are designed jointly and then, the above requirements 2)-4) can be better
satisfied. Therefore, in the following, we propose an MICF algorithm to design
$P$ OFDM pulses jointly.

\begin{figure}[t]
\begin{center}
\includegraphics[width=1\columnwidth,draft=false]{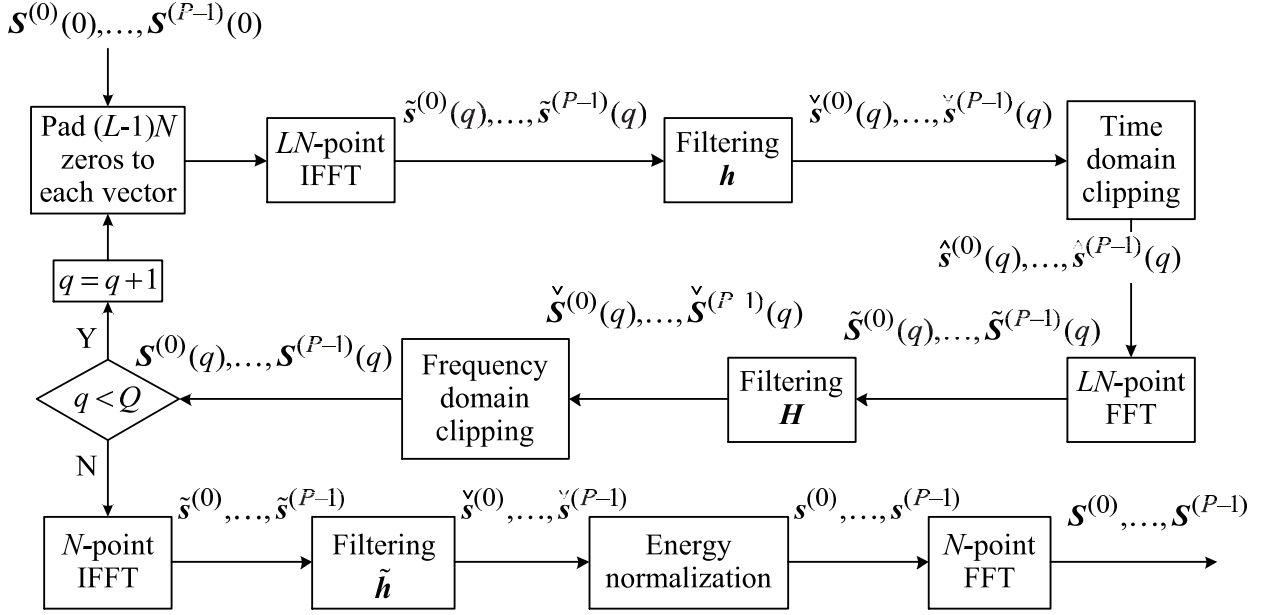}
\end{center}
\caption{Block diagram of joint multiple  OFDM sequence  design using MICF.}
\label{ICF_block}
\end{figure}

For the convenience to deal with the PAPR issue, our proposed MICF algorithm
starts with some initial random constant modular sequences
${\bS}^{(p)}(0)\in\mathbb{C}^{N\times 1}$, for $p=0,1,\ldots,P-1$. Then, at the
$q$th iteration, $\left(L-1\right)N$ zeros are padded to each sequence
${\bS}^{(p)}(q)$ as $\left[{S}_{0}^{(p)}(q),\ldots, {S}_{N-1}^{(p)}(q),\right.$
$\left. \mathbf{0}_{1\times {\left(L-1\right)N}}\right]^T$ and we obtain
$\tilde{\bs}^{(p)}(q)\in\mathbb{C}^{LN\times 1}$ by using $LN$-point IFFT, as
shown in the block diagram Fig. \ref{ICF_block}, where $\tilde{\bs}^{(p)}$
denote the time domain OFDM sequences by $L$ times over-sampling of the
continuous waveforms ${\bs}^{(p)}(t)$. Since the first ${\eta}_{max}+M-1$ and
the last ${\eta}_{max}+M-2$ samples of our desired sequences ${\bs}^{(p)}$
should be equal to zero, we apply the following time domain filter to the
sequences $\tilde{\bs}^{(p)}(q)$:
\begin{equation}
h(n)=\left\{\begin{split}
0,&\ 0\leq n\leq L({\eta}_{max}+M-1)-1\\
1,&\ L({\eta}_{max}+M-1)\leq n\leq L(N-{\eta}_{max}-M+2)-1\\
0,&\ L(N-{\eta}_{max}-M+2)\leq n\leq LN-1
\end{split}\right.,
\end{equation}
to obtain a new sequences $\check{\bs}^{(p)}(q)=
\left[\check{s}_{0}^{(p)}\left(q\right),\ldots,
\check{s}_{LN-1}^{(p)}\left(q\right)\right]^T$, where
$\check{s}_{n}^{(p)}\left(q\right)= \tilde{s}_{n}^{(p)}\left(q\right)h(n)$,
$n=0,1,\ldots,LN-1$. The time domain clipping \cite{TxzAPLOFDMSAR} is then
applied to the segment of the non-zero elements of the sequence
$\check{\bs}^{(p)}(q)$ with a pre-set constant lower bound PAPR$_d$ for a
desired PAPR, and we obtain the sequence $\hat{\bs}^{(p)}(q)$. After the
$LN$-point FFT and frequency domain filtering, we obtain the sequences
$\tilde{\bS}^{(p)}(q)$ and $\check{\bS}^{(p)}(q)$, respectively. Notice that
the frequency domain filtering is used to constrain the out-of-band radiation
caused by the time domain filtering and clipping. To deal with the constant
transmitted energy among $N$ subcarriers of the summation for all the $P$
pulses, the following frequency domain clipping is used:
\begin{equation}
\!\!\!\!\!\!\!\!\!\!\!\!{S}_{k}^{(p)}(q+1)=\left\{\begin{split}
&\sqrt{\frac{{\mathrm{Pav}(q)}\left(1+G_f\right)}
{\mathrm{P}_{k}(q)}}\check{S}_{k}^{(p)}(q),& &\textrm{if}~\mathrm{P}_{k}(q)>
{\mathrm{Pav}(q)}\left(1+G_f\right)\\
&\sqrt{\frac{{\mathrm{Pav}(q)}\left(1-G_f\right)}
{\mathrm{P}_{k}(q)}}\check{S}_{k}^{(p)}(q),& &\textrm{if}~\mathrm{P}_{k}(q)<
{\mathrm{Pav}(q)}\left(1-G_f\right)\\
&\check{S}_{k}^{(p)}(q),&&\textrm{otherwise}
\end{split}\right.,
\end{equation}
where $0\leq k\leq N-1$, we obtain ${\bS}^{(p)}(q+1)=\left[S_{0}^{(p)}(q+1),
S_{1}^{(p)}(q+1),\ldots, S_{N-1}^{(p)}(q+1)\right]^T$, and
\[\mathrm{P}_{k}(q)=\sum\limits_{p=0}^{P-1}
\left|\check{S}_{k}^{(p)}(q)\right|^2\] and
\[\mathrm{Pav}(q)=
\frac{1}{N}\sum\limits_{k=0}^{N-1}\mathrm{P}_{k}^{(p)}(q)\] are, respectively,
the transmitted energy of the $k$th subcarrier of the summation for all the $P$
pulses and the average energy
 of $N$ subcarriers for all the $P$
pulses  within a CPI. $G_f$ is a factor that we use to control the upper and
lower bounds for $\sum\limits_{p=0}^{P-1} \left|{S}_{k}^{(p)}(q+1)\right|^2$.
Thus, the value of $\sum\limits_{p=0}^{P-1}\left|{S}_{k}^{(p)}(q+1)\right|^2$
is constrained as $\sum\limits_{p=0}^{P-1}\left|{S}_{k}^{(p)}(q+1)\right|^2\in
\left[\mathrm{Pav}(q)\left(1-G_f\right),
\mathrm{Pav}(q)\left(1+G_f\right)\right]$. A smaller $G_f$ denotes that a
closer-to-constant value
$\sum\limits_{p=0}^{P-1}\left|{S}_{k}^{(p)}(q+1)\right|^2$ can be obtained.

In Fig. \ref{ICF_block}, $Q$ is a pre-set maximum iteration number. When $q=Q$,
the iteration stops and the $N$-point IFFT will be applied to the sequence
${\bS}^{(p)}(Q)\in\mathbb{C}^{N\times 1}$ to obtain
$\tilde{\bs}^{(p)}\in\mathbb{C}^{N\times 1}$. Then, a time domain filter,
\[\tilde{h}(n)=\left\{\begin{split}
0,&\ 0\leq n\leq {\eta}_{max}+M-2\\
1,&\ {\eta}_{max}+M-1\leq n\leq N-{\eta}_{max}-M+1\\
0,&\ N-{\eta}_{max}-M+2\leq n\leq N-1
\end{split}\right.,\]
is applied to $\tilde{\bs}^{(p)}$ and we obtain  sequence $\check{\bs}^{(p)}=
\left[\check{\bs}_{0}^{(p)},\ldots, \check{\bs}_{N-1}^{(p)}\right]^T$, where
$\check{\bs}_{n}^{(p)}=\tilde{\bs}_{n}^{(p)}\tilde{h}(n)$, for
$n=0,1,\ldots,N-1$. To normalize the transmitted energy and make sure
$\sum\limits_{k=0}^{N-1}\left|S_{k}^{(p)}\right|^2=\frac{1}{\mathbb{T}P}$ for
each pulse, the normalization is applied to the sequence $\check{\bs}^{(p)}$,
i.e.,
\[{s}_{n}^{(p)}=\frac{\check{s}_{n}^{(p)}}
{\sqrt{\mathbb{T}P\sum\limits_{i={\eta}_{max}+M-1}^{N-{\eta}_{max}-M+1}
\left|\check{s}_{i}^{(p)}\right|^2}},\ n=0,1,\ldots,N-1,\] and we obtain OFDM
sequence ${\bs}^{(p)}$ that accurately
 satisfies the new zero head and tail criterion 2). Finally,  sequence ${\bS}^{(p)}$ can
be obtained by using the $N$-point FFT to ${\bs}^{(p)}$. The PAPR of the
non-zero part of ${\bs}^{(p)}$ can be obtained from ${\bS}^{(p)}$
\cite{TxzAPLOFDMSAR}. The SNR degradation factor $\xi$ in (\ref{xi}) can also
be calculated from ${\bS}^{(p)},~p=0,1,\ldots,P-1$.

As a remark to finish this section is that in radar applications,
our proposed MIMO OFDM pulse design can be done off-line and as long as
one set of $P_0$ non-zero OFDM pulses are found with the above
desired properties, it is good enough and the convergence
of the  above proposed iterative
algorithm is not very important.

%%%%%%%%%%%%%%%%%%%%%%%%%%%%%%%%%%%%%%%%%%%%%%%%%%%%%%%%%%%%%%%%%%%%%%%%%%%%%%%%%%
\section{Simulation Results}\label{Simulation}
In this section, we first study the performance of our proposed MICF OFDM
sequence/pulse design by using Monte Carlo simulations. We then study the
performance of the MIMO OFDM radar detection with our designed OFDM pulses.
From what was studied in the preceding section, $P_0$
non-zero OFDM pulses are needed to be designed.

\subsection{Performance of the MICF OFDM pulse design}
\begin{figure}[b]%
\centering \subfigure[]{
\label{PAPR_Qs} %% label for first subfigure
\includegraphics[width=3.1in]{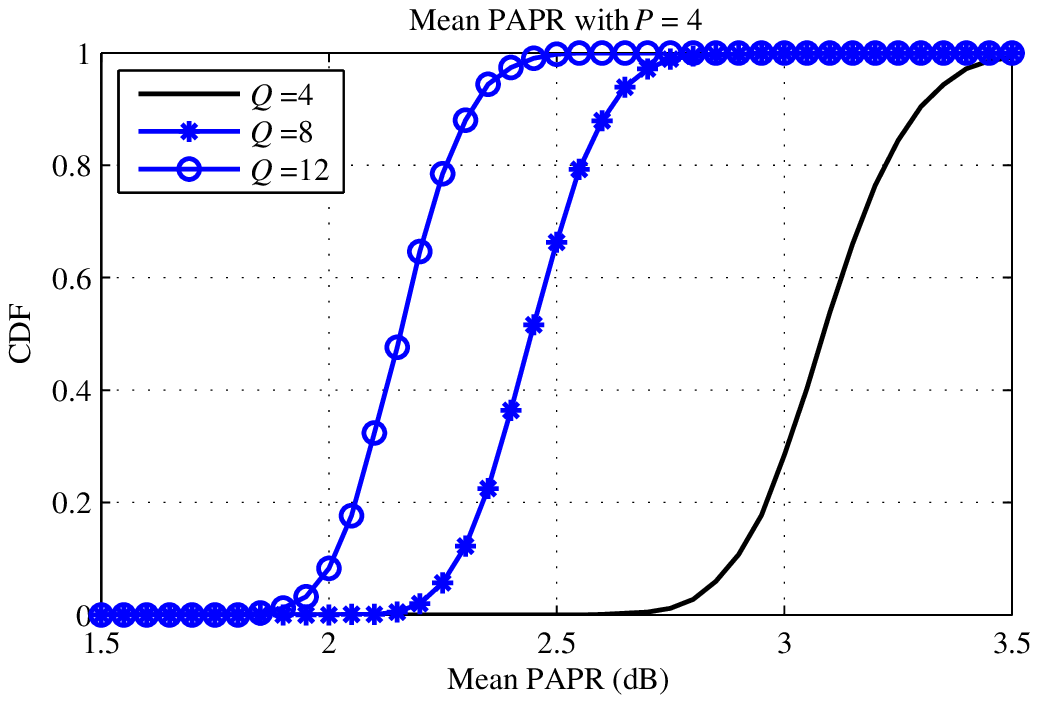}} %\hspace{0.02in}
\hspace{0.02in} \subfigure[]{ \label{SNR_Qs} %% label for second subfigure
\includegraphics[width=3.1in]{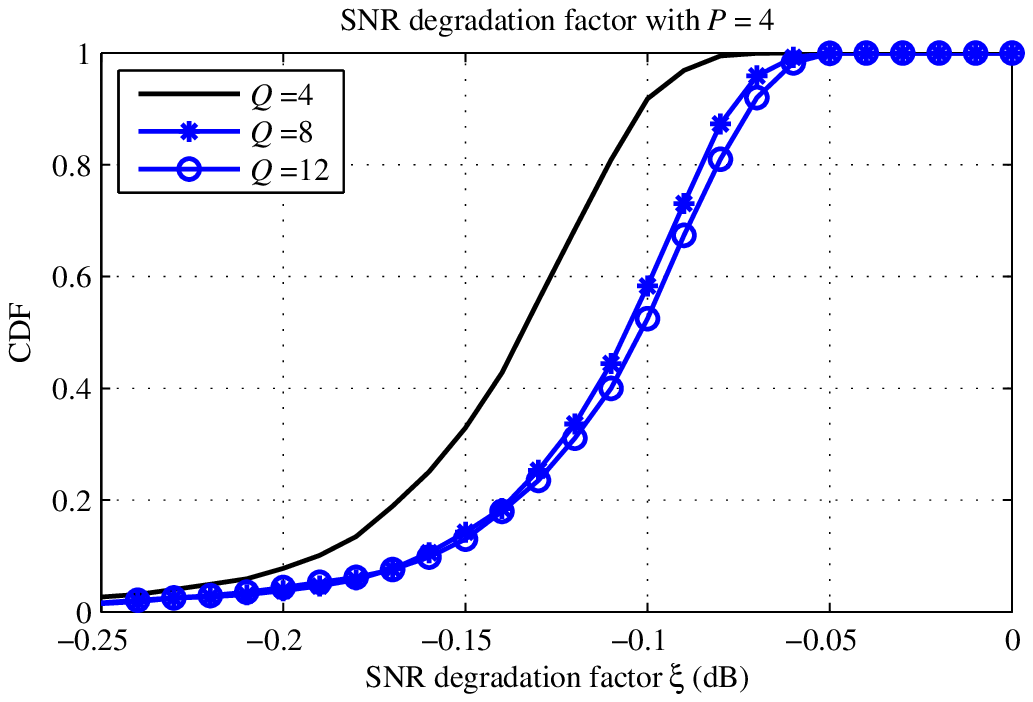}}
\caption{CDFs for different $Q$ with $P=4$, $\textrm{PAPR}_d$ $=0.1$ dB and
$G_f=10\%$: (a) Mean PAPR; (b) SNR degradation factor.}
\label{Qs} %% label for entire figure
\end{figure}

\begin{figure}[b]%
\centering \subfigure[]{
\label{PAPR_PAPRs} %% label for first subfigure
\includegraphics[width=3.1in]{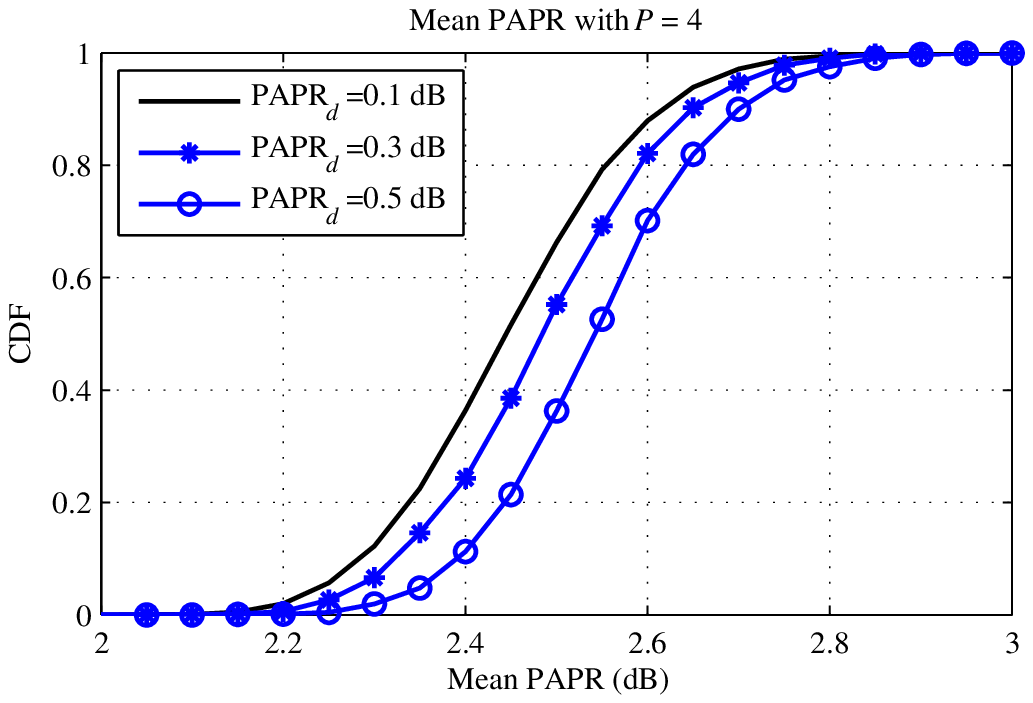}} %\hspace{0.02in}
\hspace{0.02in} \subfigure[]{ \label{SNR_PAPRs} %% label for second subfigure
\includegraphics[width=3.1in]{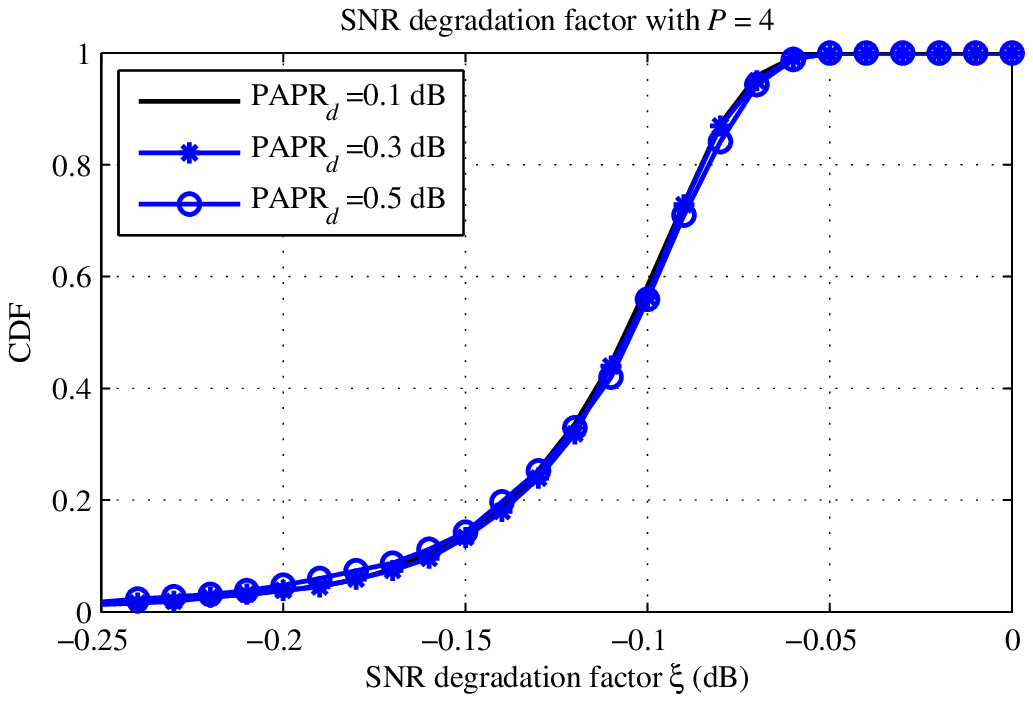}}
\caption{CDFs for different $\textrm{PAPR}_d$ with $P=4$, $Q=8$ and
$G_f=10\%$: (a) Mean PAPR; (b) SNR degradation factor.}
\label{PAPRs} %% label for entire figure
\end{figure}

\begin{figure}[t]%
\centering \subfigure[]{
\label{PAPR_Gfs} %% label for first subfigure
\includegraphics[width=3.1in]{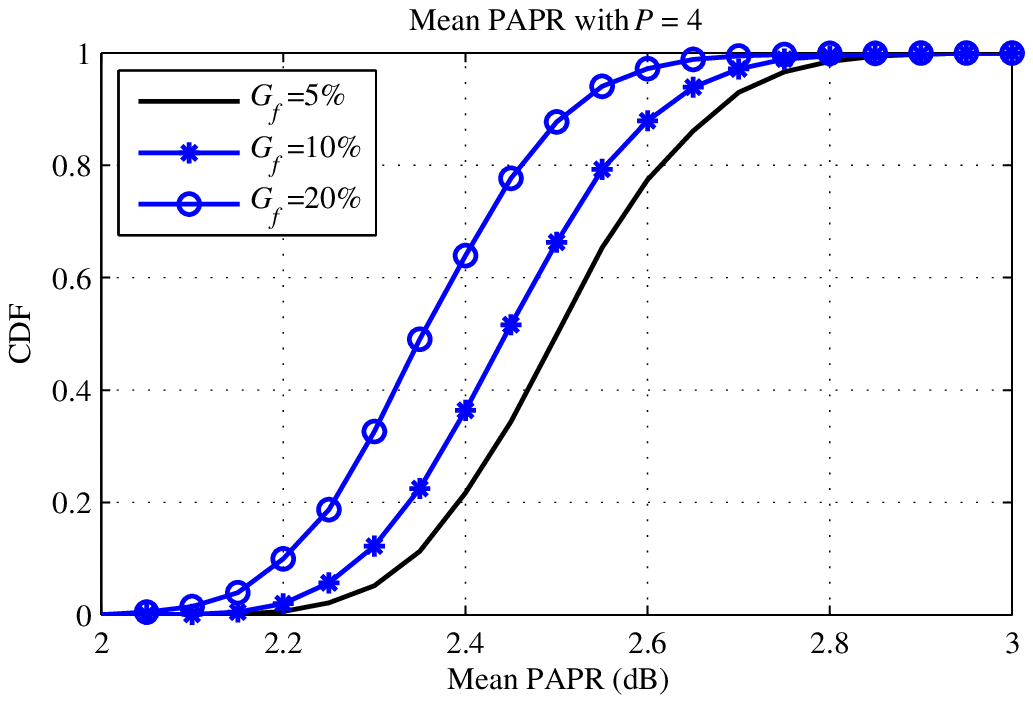}} %\hspace{0.02in}
\hspace{0.02in} \subfigure[]{ \label{SNR_Gfs} %% label for second subfigure
\includegraphics[width=3.1in]{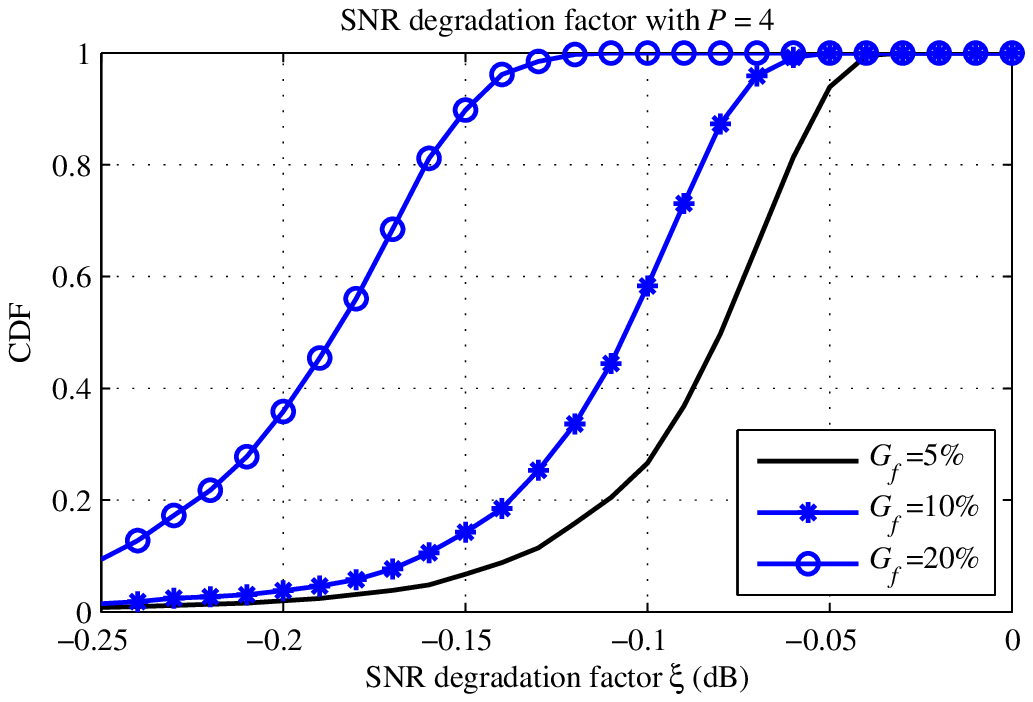}}
\caption{CDFs for different $G_f$ with $P=4$, $Q=8$ and $\textrm{PAPR}_d$
$=0.1$ dB: (a) Mean PAPR; (b) SNR degradation factor.}
\label{Gfs} %% label for entire figure
\end{figure}

\begin{figure}[b]%
\centering \subfigure[]{
\label{PAPR_Ps} %% label for first subfigure
\includegraphics[width=3.1in]{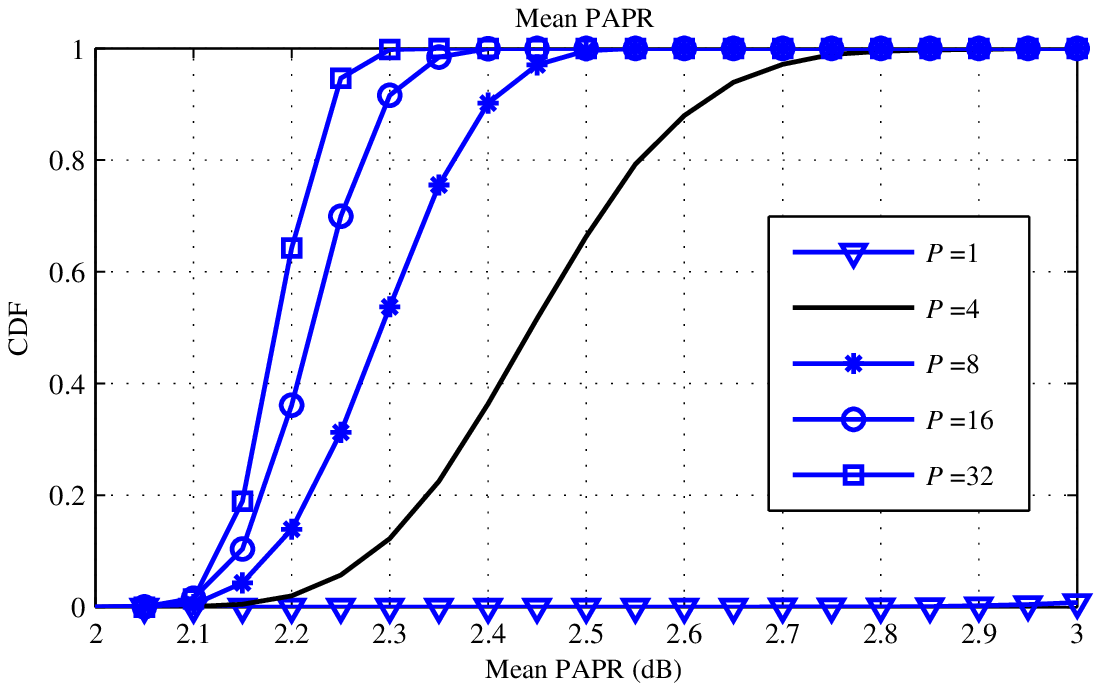}} %\hspace{0.02in}
\hspace{0.02in} \subfigure[]{ \label{SNR_Ps} %% label for second subfigure
\includegraphics[width=3.1in]{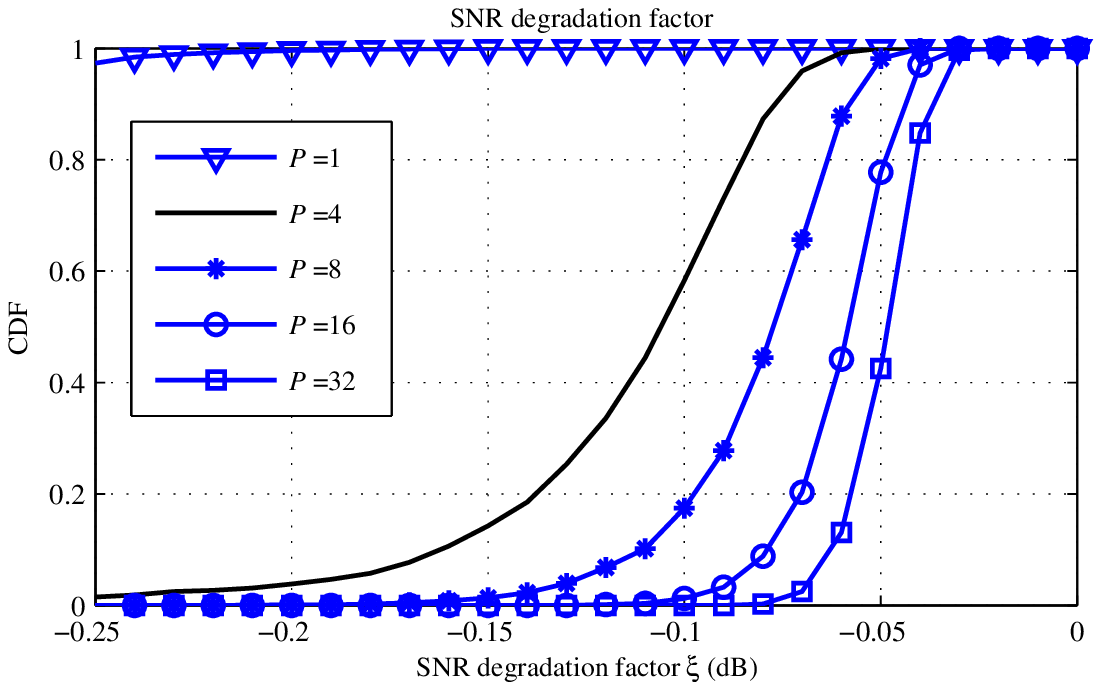}}
\caption{CDFs for different $P$ with $Q=8$, $\textrm{PAPR}_d$ $=0.1$ dB and
$G_f=10\%$: (a) Mean PAPR; (b) SNR degradation factor.}
\label{Ps} %% label for entire figure
\end{figure}

In this subsection, we first see the performance of the MICF OFDM pulse design
algorithm. We set
the number of range cells
$M=96$, the maximum relative time delays
$\eta_{max}=40$,
the number of subcarriers
$N=302$ and
the non-zero pulse length
$N_t=33$. To achieve a
sufficiently accurate PAPR estimate, we set the over-sampling ratio $L=4$
\cite{ArmstrongELPAPRreduction2002246, SeungIEEEWCPAPRoverview200556}. We
evaluate the mean PAPR of the $P$ pulses and the SNR degradation factor $\xi$
by using the standard Monte Carlo technique with $2000$ independent trials. In
each trial, the $k$th element of an initial sequence
${\bS}_{\alpha}^{(p)}{(0)}$ is set as
${S}_{\alpha,k}^{(p)}(0)={\frac{1}{\sqrt{N\mathbb{T}P}}}e^{j2\pi
\varphi_k^{(p)}},\ k=0,1,\ldots,N-1$, where $\varphi_k^{(p)}$ is uniformly
distributed within the interval $[0,2\pi]$.  In Figs. \ref{Qs}-\ref{Gfs}, we
plot the cumulative distribution functions (CDF) of the mean PAPR and the SNR
degradation factor $\xi$ with $P=4$. The curves in Fig. \ref{Qs} denote that,
with the increase of the maximum iteration number $Q$, the mean PAPR decreases
and  $\xi$ increases. Therefore, better $P$ OFDM pulses with
lower mean PAPR and larger $\xi$ can be obtained by using a larger iteration
number $Q$. The curves in Fig. \ref{PAPRs} show that, with the increase of
$\textrm{PAPR}_d$, the mean PAPR increases and  $\xi$ decreases, in the
meantime the mean PAPR change is more sensitive than the change of $\xi$ for
different $\textrm{PAPR}_d$. Similarly, the curves in Fig. \ref{Gfs} indicate
that the mean PAPR is decreased and the SNR degradation is increased, when
$G_f$ is increased. In summary, the simulation results of mean PAPR and $\xi$
are better than the corresponding results for single OFDM pulse design
(corresponding to the case of $P=1$)
in
\cite{TxzAPLOFDMSAR} even though with a small value of $Q$
as shown in Fig.  \ref{Ps}, which is
because the joint design of
$P$ OFDM pulses  provides more degrees of
freedom for the MICF algorithm. We
also plot the CDFs of mean PAPR and $\xi$ for different pulse numbers
 $P$ with
$Q=8$, $\textrm{PAPR}_d$ $=0.1$ dB and $G_f=10\%$ in Fig. \ref{Ps}. The curves
in Fig. \ref{Ps} show  that, with the increase of $P$, the mean PAPR and $\xi$
are significantly improved,
where one can see that the single OFDM pulse design, i.e.,
when $P=1$, is poor due to the small iteration number $Q=8$ is
used. It further indicates the benefits of the proposed
MICF algorithm with
joint design of
 $P$ OFDM pulses.

According to the above analysis, the mean PAPR and $\xi$ are interacting each
other. In practice, it is necessary to consider the constraints of both mean
PAPR and $\xi$ at the same time. In Table \ref{RunNumbers}, we count the
numbers of trials under the conditions of $\xi\geq -0.08$ \textnormal{dB} and
mean PAPR $\leq 2.2$ within the $2000$ Monte Carlo independent trials for
$Q=8$, $\textrm{PAPR}_d$ $=0.1$ dB and $G_f=10\%$. The numbers of trials are
increased significantly with the increase of $P$. According to our simulations,
there are $7$ trials that satisfy the conditions of $\xi\geq -0.04$ dB and mean
PAPR $\leq 2.1$ dB with $P=32$, which is not shown in
Table  \ref{RunNumbers}.

\begin{table}[!htp]
\caption{Numbers of Monte Carlo trials for $\xi\geq -0.08$ \textnormal{dB} and
mean PAPR $\leq 2.2$ \textnormal{dB} with $Q=8$, $\textrm{PAPR}_d$ $=0.1$
\textnormal{dB} and $G_f=10\%$} \label{RunNumbers}
\begin{center}
\begin{tabular}{|c|c|c|c|}
\hline  %
$P=4$  & $P=8$  & $P=16$  & $P=32$\\
\hline
14   & 169     & 680 & 1282  \\
\hline
\multicolumn{4}{|c|}{Total number of trials: 2000} \\
\hline
\end{tabular}
\end{center}
\end{table}

\subsection{Performance of the MIMO OFDM radar range reconstruction}
\begin{figure}[b]%
\begin{center}
\includegraphics[width=1\columnwidth,draft=false]{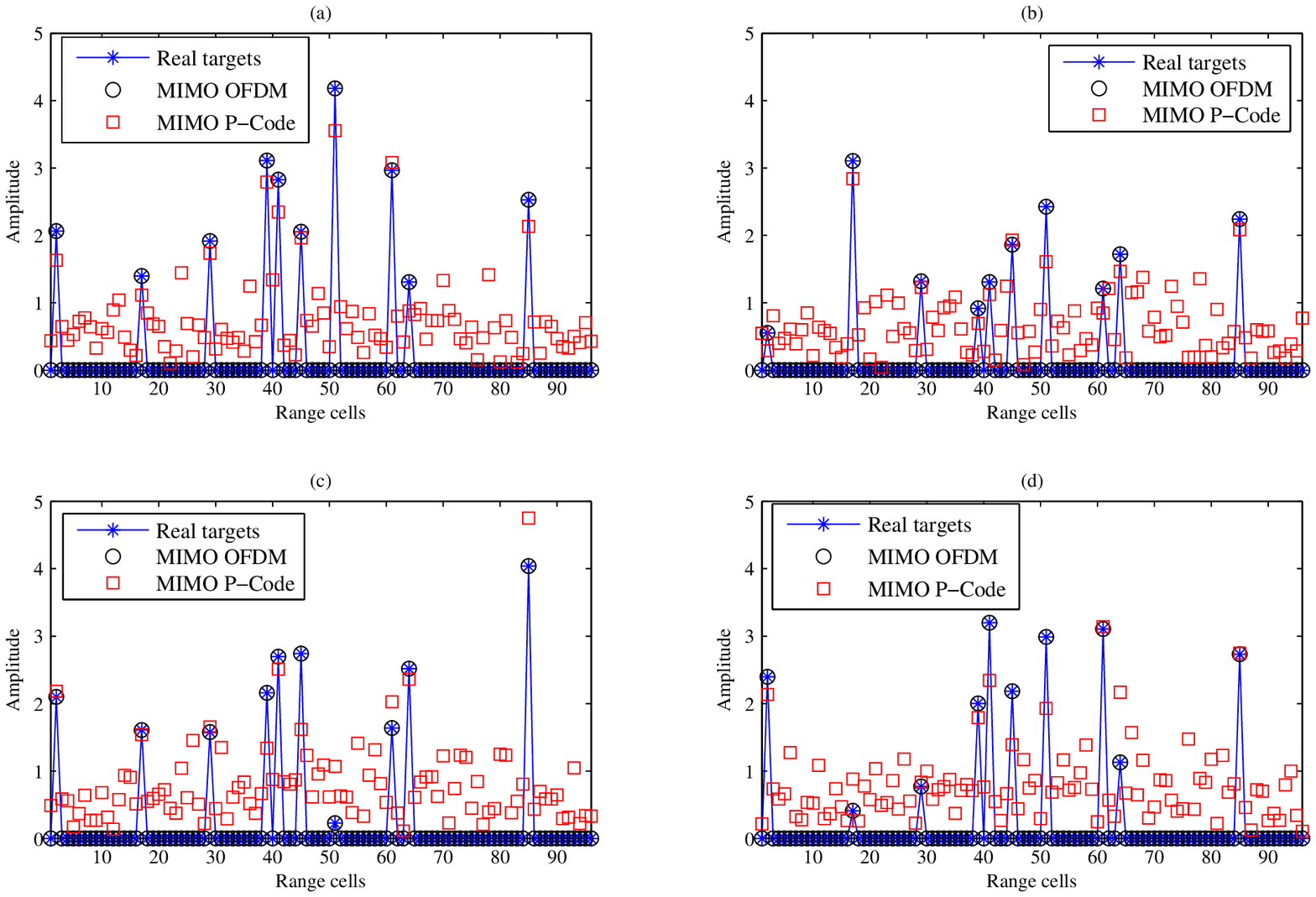}
\end{center}
\caption{Amplitudes of targets for different transmitter and receiver pairs
after the range reconstructions without noise using polyphase waveforms and our
designed OFDM pulses with transmitter and receiver pair: (a) $(\alpha,
\beta)=(1,1)$; (b)
 $(\alpha, \beta)=(2,1)$; (c)
 $(\alpha, \beta)=(1,2)$;
 (d) $(\alpha, \beta)=(2,2)$.}
\label{Amplitude_No_noise} %% label for entire figure
\end{figure}

\begin{figure}[b]%
\begin{center}
\includegraphics[width=1\columnwidth,draft=false]{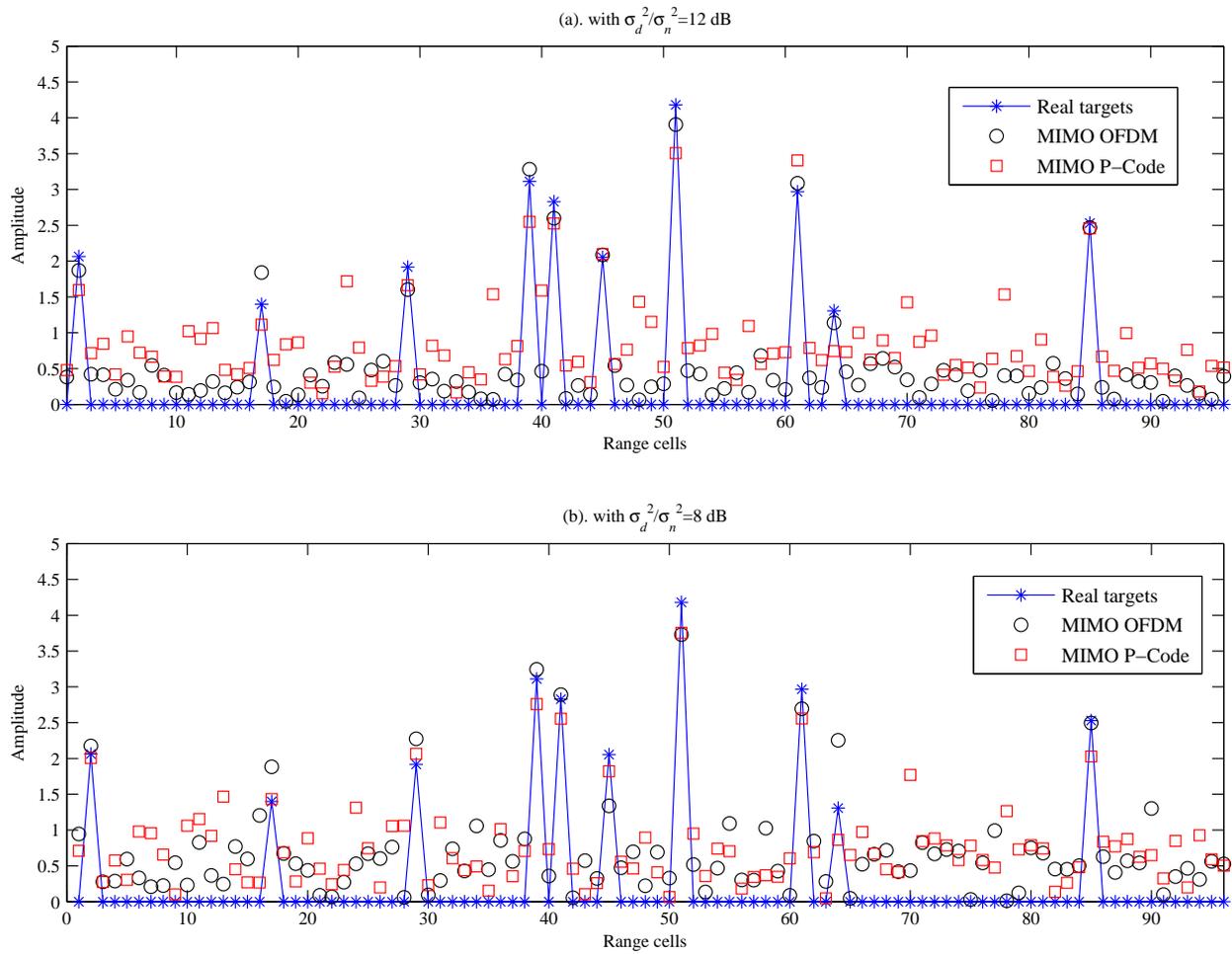}
\end{center}
\caption{Amplitudes of targets for
 $(\alpha, \beta)=(1,1)$  after
the range reconstructions using polyphase waveforms and our designed OFDM pulses:
(a) with $\frac{\sigma_d^2}{\sigma_n^2}=12$ dB; (b) with
$\frac{\sigma_d^2}{\sigma_n^2}=8$ dB.}
\label{MIMO_Code_SNRs} %% label for entire figure
\end{figure}

In this subsection, we investigate the performance of the MIMO OFDM radar range
reconstruction. We set the
bandwidth
$B=150$ MHz,
the carrier frequency
$f_c=9$ GHz,
the number of range cells $M=96$,
the maximum  relative time delay
$\eta_{max}=40$, the number of subcarriers
$N=309$,
the length of a non-zero pulse
$N_t=40$,
the number of transmitters $\mathbb{T}=2$ and
the number of receivers $\mathbb{R}=2$, the number of pulses
 $P=2$. We  use
our
designed OFDM pulses with the degradation factor
$\xi=-0.07$ dB and mean PAPR $=2.06$ dB. For convenience, the time delays
$\eta_{\beta,\alpha}$ are randomly chosen
 within the integer  interval
$\left[0,\eta_{max}\right]$ as $\eta_{1,1}=17$,
$\eta_{1,2}=0$, $\eta_{2,1}=6$,
$\eta_{2,2}=32$. Considering a single range line, the
targets (non-zero RCS coefficients) are included in $10$ random range cells
located from $10000$ m to $10096$ m. The RCS coefficients
$g_{\beta,\alpha,m}$ within the $10$ range cells are independent and
obey complex white Gaussian distribution with zero-mean and variance
$\sigma_d^2$, i.e.,
$g_{\beta,\alpha,m}\sim\mathcal{CN}\left(0,\sigma_d^2\right)$ for
all receivers $\beta$ and transmitters $\alpha$. For comparison, we
also use the first two polyphase waveforms of the polyphase code set with
length $40$ in \cite{HaiDengPolyphaseCodeIEEETSP20043126}. The two polyphase
waveforms are applied in the two transmitters, respectively. After pulse
compression with matched filtering and pulse coherent integration, the range
reconstruction results are shown in Figs.
\ref{Amplitude_No_noise}-\ref{MIMO_Code_SNRs} with red square marks that are
denoted as ``MIMO P-Code.'' For the  better display, in this and following
simulations, the pulse compression and integration gains of all the range
reconstruction results are normalized.

In Fig. \ref{Amplitude_No_noise}, we plot the range reconstruction results of
all the transmitter and receiver pairs with $\sigma_d^2=1$ and without noise.
Comparing with the real target amplitudes (with blue solid line with asterisk
marks), the results show that the MIMO OFDM range reconstruction is precise for
all the transmitter and receiver pairs. It also indicates that there is no any
interference between different transmitters and the full spatial diversity can
be achieved by using our proposed MIMO OFDM radar. Meanwhile, the benefit of
the IRCI free range reconstruction by using CP based OFDM radar  still holds.
 However, because of the non-zero cross correlation (or
non-orthogonality) between the two polyphase waveforms as well as the range
sidelobes of the autocorrelation functions, some targets can not be
reconstructed correctly as shown in Fig. \ref{Amplitude_No_noise},
and thus, the
spatial diversity can not be clearly obtained by using the polyphase waveforms.
Moreover, the range reconstruction results of some range cells without target
by using the polyphase waveforms are much larger than $0$. We also consider the
range reconstruction performances for\footnote{Notice that, according to
(\ref{OLSNRrtm}) and normalized transmitted energy constraint, the SNR of the
received signals are about $-10.04$ dB and $-14.04$ dB for
$\frac{\sigma_d^2}{\sigma_n^2}=12$ dB and $8$ dB, respectively.}
$\frac{\sigma_d^2}{\sigma_n^2}=12$ dB and $8$ dB, and the simulation results
for the transmitter and receiver pair $(\alpha, \beta)=(1,1)$ are plotted in Fig.
\ref{MIMO_Code_SNRs}. The results show that the performances of our proposed
MIMO OFDM radar are better than
 that by using the polyphase waveforms, especially
for a larger SNR, for example, when $\frac{\sigma_d^2}{\sigma_n^2}=12$ dB.

\begin{figure}[b]%
\begin{center}
\includegraphics[width=1\columnwidth,draft=false]{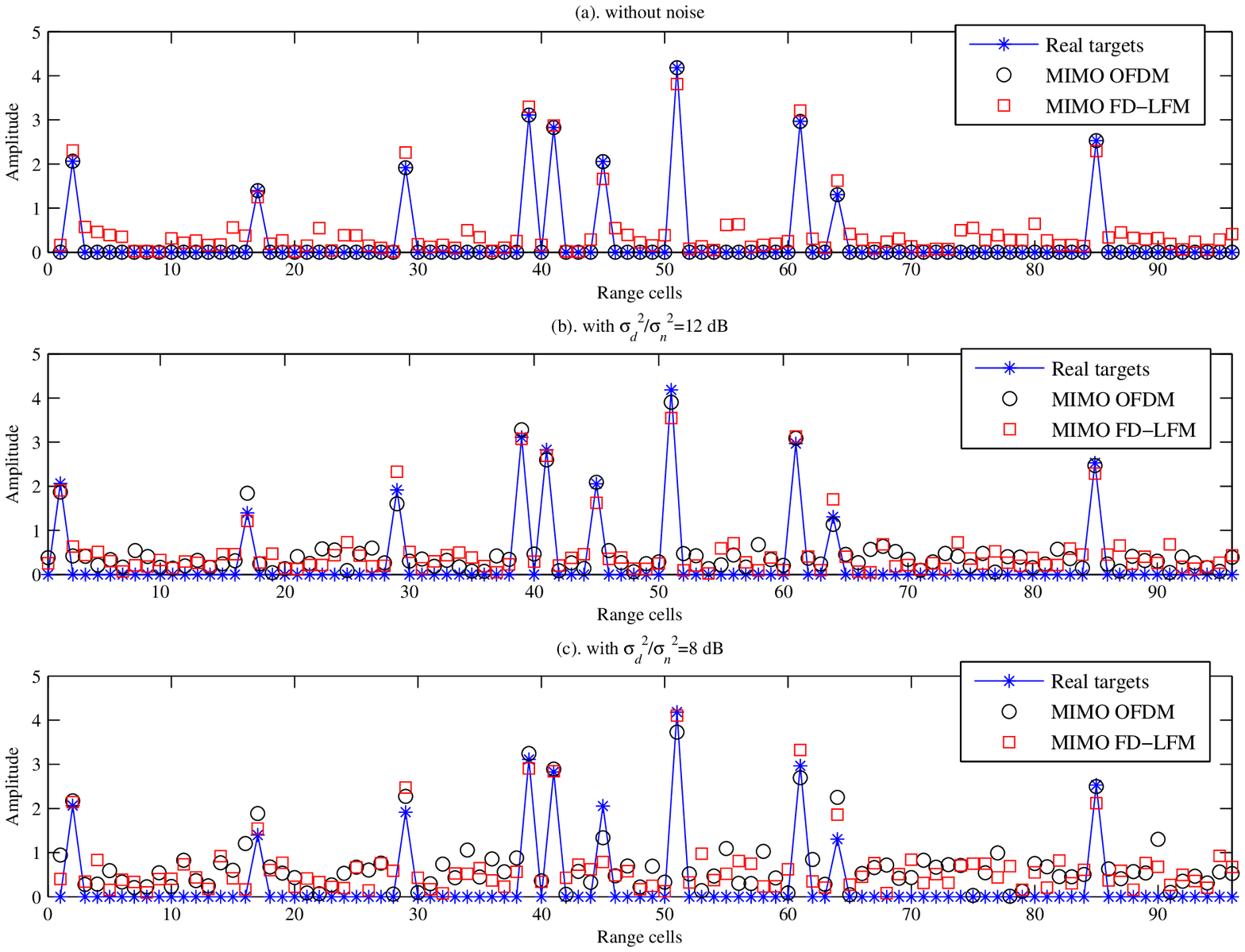}
\end{center}
\caption{Amplitudes of targets for
 $(\alpha, \beta)=(1,1)$  after
the range reconstructions using the frequency division LFM waveforms and
our designed OFDM pulses: (a) without noise; (b) with
$\frac{\sigma_d^2}{\sigma_n^2}=12$ dB; (c) with
$\frac{\sigma_d^2}{\sigma_n^2}=8$ dB.}
\label{MIMO_FD_LFM_SNRs} %% label for entire figure
\end{figure}

For further comparison, we also consider the frequency division MIMO radar, in
which each transmitted waveform is assigned an independent and non-overlapped
frequency band with bandwidth $B$. Thus, the orthogonality of the transmitted
waveforms is guaranteed in this radar system despite
time delays, but a $\mathbb{T}$
times more bandwidth (i.e., $\mathbb{T}B$) is required. By using LFM waveforms and
the above simulation parameters, we obtain and plot the range reconstruction
results in Fig. \ref{MIMO_FD_LFM_SNRs} with red square marks that are denoted
as ``MIMO FD-LFM.'' By comparing with the true target amplitudes, the results
indicate that the performances of  our proposed MIMO OFDM radar
are obviously better than
 the ``MIMO FD-LFM'' radar for the cases without noise and
$\frac{\sigma_d^2}{\sigma_n^2}=12$ dB. It is because that the IRCI across
the range cells occurs by using LFM waveforms, even through
 the cross correlation
can be completely avoided by using frequency division. The performances of
``MIMO OFDM'' and ``MIMO FD-LFM'' are similar to each other for
$\frac{\sigma_d^2}{\sigma_n^2}=8$ dB. However, in ``MIMO FD-LFM'' the bandwidth
requirement is $300$ MHz, twice more. We believe that the IRCI will be more
serious by using LFM waveforms when more range cells are included in targets,
and the benefit of our proposed MIMO OFDM radar will be more obvious.

\section{Conclusion}\label{Conclusion}
In this paper, we proposed a novel  frequency-band
 shared  and sufficient CP based MIMO
OFDM radar
range reconstruction method by using our newly proposed and designed
MIMO OFDM pulses that are in the same frequency band but
orthogonal each other for every subcarrier  in the discrete frequency domain.
This range reconstruction algorithm with the orthogonality
of the MIMO OFDM signals  can provide the advantage of
avoiding the interference  between different transmitters,
even when there are time delays among the signals from different transmitters,
and
achieving the full
spatial diversity. Meanwhile, due to
the sufficient CP insertion to each pulse with the zero head and tail values
in the discrete time domain,
the range reconstruction is IRCI free and
the proposed system does not have  the energy redundancy.
Our proposed range reconstruction is a joint pulse compression and
pulse coherent integration, after which
 the SNR
was analyzed.
We then  proposed four design criteria for multiple OFDM pulses.
To achieve the orthogonality for every subcarrier
 in the discrete frequency domain
across multiple transmitters, complex orthogonal designs
were adopted, with which only non-zero-valued OFDM pulses for
the first transmitter are needed to be designed.
To maximize the SNR,
 a closed-form solution was proposed by using the paraunitary filterbank
theory.
Considering the trade-off between the PAPR and the SNR degradation within the
range reconstruction, we also proposed an
MICF joint OFDM pulse design method
to obtain OFDM pulses with low PAPRs and insignificant SNR degradation. We
finally presented
 some simulations to demonstrate the performance of the proposed
OFDM pulse design method. By comparing with the frequency-band
 shared MIMO radar
using polyphase code waveforms and frequency division MIMO radar using LFM
waveforms, we provided some simulations to illustrate the advantage, such as
the full spatial diversity and free IRCI,  after the range reconstruction, of
the proposed MIMO OFDM radar.

This paper provides a framework on frequency-band shared
statistical MIMO OFDM radar with IRCI free
and inter-transmitter-interference (ITI) free range reconstruction.
Some interesting research problems remain. One of them would be
on how to deal with the trade-off between the non-zero pulse number $P_0$
and the total pulse number $P$ in a CPI. The other one would be
on how to search the parameters in the paraunitary matrix to satisfy the
ideal flat spectral power criterion 3) and also have good PAPR property, i.e.,
satisfy criterion 4).

As a final remark, this paper only considers statistial
MIMO radar where multiple OFDM pulses with sufficient CP are transmitted
by each transmitter in a coherent processing interval (CPI).
Colocated MIMO OFDM radar has been recently considered
in \cite{yhcao} where only one OFDM pulse with sufficient CP
is transmitted in a CPI at each transmitter.

%\bibliographystyle{IEEEtran}%unsrt  %
%\bibliography{IEEEabrv,D:/assignment/Report/Paper/MyReference}
\end{document}